\documentclass[twocolumn]{aastex63}

\usepackage{color}
\usepackage{graphicx}
\usepackage{amsmath, amssymb, bm}
\usepackage{CJK}

\definecolor{red}{rgb}{1.0,0.0,0.0}

\newcommand{\Lp}{$L$'}

\newcommand{\uatnum}[1]{\href{http://vocabs.ands.org.au/repository/api/lda/aas/the-unified-astronomy-thesaurus/current/resource.html?uri=http://astrothesaurus.org/uat/#1}{#1}}

\turnoffedit

\submitjournal{AAS Journals}

\shorttitle{PDS 70 Vortex Imaging}
\shortauthors{Wang et al.}

\begin{document}
\begin{CJK*}{UTF8}{gbsn}

\title{Keck/NIRC2 \Lp{}-Band Imaging of Jovian-Mass Accreting Protoplanets around PDS 70}

\correspondingauthor{Jason Wang}
\email{jwang4@caltech.edu}

\author[0000-0003-0774-6502]{Jason J. Wang (王劲飞)}
\altaffiliation{51 Pegasi b Fellow}
\affiliation{Department of Astronomy, California Institute of Technology, Pasadena, CA 91125, USA}

\author{Sivan Ginzburg}
\altaffiliation{51 Pegasi b Fellow}
\affiliation{Department of Astronomy, University of California at Berkeley, CA 94720, USA}

\author[0000-0003-1698-9696]{Bin Ren (任彬)}
\affiliation{Department of Astronomy, California Institute of Technology, Pasadena, CA 91125, USA}

\author[0000-0003-0354-0187]{Nicole Wallack}
\affiliation{Division of Geological \& Planetary Sciences, California Institute of Technology, Pasadena, CA 91125, USA}

\author[0000-0002-8518-9601]{Peter Gao}
\altaffiliation{51 Pegasi b Fellow}
\affiliation{Department of Astronomy, University of California at Berkeley, CA 94720, USA}

\author[0000-0002-8895-4735]{Dimitri Mawet}
\affiliation{Department of Astronomy, California Institute of Technology, Pasadena, CA 91125, USA}
\affiliation{Jet Propulsion Laboratory, California Institute of Technology, 4800 Oak Grove Dr.,Pasadena, CA 91109, USA}

\author{Charlotte Z. Bond}
\affiliation{Institute for Astronomy, University of Hawaii, 2680 Woodlawn Drive, Honolulu, HI 96822, USA}
\affiliation{W. M. Keck Observatory, 65-1120 Mamalahoa Hwy, Kamuela, HI, USA}

\author{Sylvain Cetre}
\affiliation{W. M. Keck Observatory, 65-1120 Mamalahoa Hwy, Kamuela, HI, USA}

\author{Peter Wizinowich}
\affiliation{W. M. Keck Observatory, 65-1120 Mamalahoa Hwy, Kamuela, HI, USA}

\author[0000-0002-4918-0247]{Robert J. De Rosa}
\affiliation{European Southern Observatory, Alonso de Cordova 3107, Vitacura, Santiago, Chile}

\author[0000-0003-4769-1665]{Garreth Ruane}
\affiliation{Jet Propulsion Laboratory, California Institute of Technology, 4800 Oak Grove Dr.,Pasadena, CA 91109, USA}

\author[0000-0003-2232-7664]{Michael C. Liu}
\affiliation{Institute for Astronomy, University of Hawaii, 2680 Woodlawn Drive, Honolulu, HI 96822, USA}


\author{Olivier Absil}
\affiliation{Space sciences, Technologies \& Astrophysics Research (STAR) Institute, University of Li\`ege, Li\`ege, Belgium}

\author{Carlos Alvarez}
\affiliation{W. M. Keck Observatory, 65-1120 Mamalahoa Hwy, Kamuela, HI, USA}

\author[0000-0002-1917-9157]{Christoph Baranec}
\affiliation{Institute for Astronomy, University of Hawai`i at M\={a}noa, 640 North A`ohoku Place, Hilo, HI 96720-2700, USA}

\author[0000-0002-9173-0740]{\'Elodie Choquet}
\affiliation{Aix Marseille Univ, CNRS, CNES, LAM, Marseille, France}

\author{Mark Chun}
\affiliation{Institute for Astronomy, University of Hawai`i at M\={a}noa, 640 North A`ohoku Place, Hilo, HI 96720-2700, USA}

\author{Denis Defr\`ere}
\affiliation{Space sciences, Technologies \& Astrophysics Research (STAR) Institute, University of Li\`ege, Li\`ege, Belgium}

\author{Jacques-Robert Delorme}
\affiliation{Department of Astronomy, California Institute of Technology, Pasadena, CA 91125, USA}

\author[0000-0002-5092-6464]{Gaspard Duch\^ene}
\affiliation{Department of Astronomy, University of California at Berkeley, CA 94720, USA}
\affiliation{Universit\'e Grenoble-Alpes, CNRS Institut de Plan\'etologie et d'Astrophysique (IPAG), F-38000 Grenoble, France}

\author{Pontus Forsberg}
\affiliation{Department of Materials Science and Engineering, \AA ngstr\"om Laboratory, Uppsala University, Box 534, 751 21, Uppsala, Sweden}

\author[0000-0003-3230-5055]{Andrea Ghez}
\affiliation{Department of Physics \& Astronomy, 430 Portola Plaza, University of California, Los Angeles, CA 90095, USA}

\author[0000-0002-1097-9908]{Olivier Guyon}
\affiliation{Subaru Telescope, National Astronomical Observatory of Japan, 650 North Aohoku Place, Hilo, HI 96720, USA}
\affiliation{Steward Observatory, University of Arizona, Tucson, AZ 85721, USA}
\affiliation{Astrobiology Center of NINS, 2-21-1 Osawa, Mitaka, Tokyo 181-8588, Japan}

\author{Donald N. B. Hall}
\affiliation{Institute for Astronomy, University of Hawai`i at M\={a}noa, 640 North A`ohoku Place, Hilo, HI 96720-2700, USA}

\author{Elsa Huby}
\affiliation{LESIA, Observatoire de Paris, Universit\'e PSL, CNRS, Sorbonne Universit\'e, Universit\'e de Paris, 5 place Jules Janssen, 92195 Meudon, France}

\author{A\"issa Jolivet}
\affiliation{Space sciences, Technologies \& Astrophysics Research (STAR) Institute, University of Li\`ege, Li\`ege, Belgium}

\author[0000-0003-0054-2953]{Rebecca Jensen-Clem}
\affiliation{Department of Astronomy \& Astrophysics, University of California, Santa Cruz, CA95064, USA}

\author[0000-0001-5213-6207]{Nemanja Jovanovic}
\affiliation{Department of Astronomy, California Institute of Technology, Pasadena, CA 91125, USA}

\author{Mikael Karlsson}
\affiliation{Department of Materials Science and Engineering, \AA ngstr\"om Laboratory, Uppsala University, Box 534, 751 21, Uppsala, Sweden}

\author{Scott Lilley}
\affiliation{W. M. Keck Observatory, 65-1120 Mamalahoa Hwy, Kamuela, HI, USA}

\author{Keith Matthews}
\affiliation{Department of Astronomy, California Institute of Technology, Pasadena, CA 91125, USA}

\author[0000-0002-1637-7393]{Fran\c{c}ois M\'enard} 
\affiliation{Universit\'e Grenoble-Alpes, CNRS Institut de Plan\'etologie et d'Astrophysique (IPAG), F-38000 Grenoble, France}

\author[0000-0001-6126-2467]{Tiffany Meshkat}
\affiliation{IPAC, California Institute of Technology, M/C 100-22, 1200 East California Boulevard, Pasadena, CA 91125, USA}

\author[0000-0001-6205-9233]{Maxwell Millar-Blanchaer}
\affiliation{Department of Astronomy, California Institute of Technology, Pasadena, CA 91125, USA}
\affiliation{Jet Propulsion Laboratory, California Institute of Technology, 4800 Oak Grove Dr.,Pasadena, CA 91109, USA}

\author[0000-0001-5172-4859]{Henry Ngo}
\affiliation{NRC Herzberg Astronomy and Astrophysics, 5071 West Saanich Road, Victoria, British Columbia, Canada}

\author{Gilles Orban de Xivry}
\affiliation{Space sciences, Technologies \& Astrophysics Research (STAR) Institute, University of Li\`ege, Li\`ege, Belgium}

\author[0000-0001-5907-5179]{Christophe Pinte} 
\affiliation{Monash Centre for Astrophysics (MoCA) and School of Physics and Astronomy, Monash University, Clayton Vic 3800, Australia}
\affiliation{Universit\'e Grenoble-Alpes, CNRS Institut de Plan\'etologie et d'Astrophysique (IPAG), F-38000 Grenoble, France}

\author{Sam Ragland}
\affiliation{W. M. Keck Observatory, 65-1120 Mamalahoa Hwy, Kamuela, HI, USA}

\author{Eugene Serabyn}
\affiliation{Jet Propulsion Laboratory, California Institute of Technology, 4800 Oak Grove Dr.,Pasadena, CA 91109, USA}

\author{Ernesto Vargas Catal\'an}
\affiliation{Department of Materials Science and Engineering, \AA ngstr\"om Laboratory, Uppsala University, Box 534, 751 21, Uppsala, Sweden}

\author[0000-0002-4361-8885]{Ji Wang}
\affiliation{Department of Astronomy, The Ohio State University,100 W 18th Ave, Columbus, OH 43210, USA}

\author{Ed Wetherell}
\affiliation{W. M. Keck Observatory, 65-1120 Mamalahoa Hwy, Kamuela, HI, USA}

\author[0000-0001-5058-695X]{Jonathan P. Williams}
\affiliation{Institute for Astronomy, University of Hawaii, 2680 Woodlawn Drive, Honolulu, HI 96822, USA}

\author[0000-0001-7591-2731]{Marie Ygouf}
\affiliation{NASA Exoplanet Science Institute, IPAC, Pasadena, CA 91125, USA}

\author{Ben Zuckerman}
\affiliation{Department of Physics \& Astronomy, 430 Portola Plaza, University of California, Los Angeles, CA 90095, USA}

\begin{abstract}

We present \Lp{}-band imaging of the PDS 70 planetary system with Keck/NIRC2 using the new infrared pyramid wavefront sensor. We detected both PDS 70 b and c in our images, as well as the front rim of the circumstellar disk. After subtracting off a model of the disk, we measured the astrometry and photometry of both planets. Placing priors based on the dynamics of the system, we estimated PDS 70 b to have a semi-major axis of $20^{+3}_{-4}$~au and PDS 70 c to have a semi-major axis of $34^{+12}_{-6}$~au (95\% credible interval). We fit the spectral energy distribution (SED) of both planets. For PDS 70 b, we were able to place better constraints on the red half of its SED than previous studies and inferred the radius of the photosphere to be 2-3~$R_{Jup}$. The SED of PDS 70 c is less well constrained, with a range of total luminosities spanning an order of magnitude. With our inferred radii and luminosities, we used evolutionary models of accreting protoplanets to derive a mass of PDS 70 b between 2 and 4 $M_{\textrm{Jup}}$ and a mean mass accretion rate between $3 \times 10^{-7}$ and $8 \times 10^{-7}~M_{\textrm{Jup}}/\textrm{yr}$. For PDS 70 c, we computed a mass between 1 and 3 $M_{\textrm{Jup}}$ and mean mass accretion rate between $1 \times 10^{-7}$ and $5 \times~10^{-7} M_{\textrm{Jup}}/\textrm{yr}$. The mass accretion rates imply dust accretion timescales short enough to hide strong molecular absorption features in both planets' SEDs.

\end{abstract}

\keywords{Exoplanet formation (\uatnum{492}), Exoplanet atmospheres (\uatnum{487}), Orbit determination (\uatnum{1175}), Exoplanet dynamics (\uatnum{490}), Coronagraphic imaging (\uatnum{313})}

\section{Introduction} \label{sec:intro}
Planet formation is a difficult process to study directly. The two primary channels to form giant planets from circumstellar material are thought to be core accretion \citep{Pollack1996} and disk instability \citep{Bodenheimer1974,Boss1998}. Disk instability forms planets within $10^{5}$~yr \citep{Boss1998}, and core accretion takes a few Myr \citep{Pollack1996, Piso2014, Piso2015}. We can look at relatively young planets ($\sim$10-100~Myr) for clues of how they formed, as their formation history is encoded in the residual heat radiating from them \citep{Baraffe2003, Marley2007}. However, the predicted luminosity of cooling young planets may be degenerate between formation channels \citep{Mordasini2017}, so it is not a replacement for observing planet formation directly.

Because of the relatively short timescales for planet formation and the paucity of nearby ($\lesssim$200 pc), young ($\lesssim$10 Myr) stars around which we can detect young forming planets on Solar System scales, capturing a planet in the process of forming is challenging. Even for systems that are at favorable ages and distances for direct imaging, it is difficult to distinguish forming planets from circumstellar dust that can appear clumpy or are shrouding the planets. In both the HD 100546 and LkCa 15 systems, there have been reported detections of still-forming protoplanets \citep{Kraus2012, Quanz2013, Currie2015,Sallum2015}, but other studies have found these signals to be consistent with dust emission \citep{Thalmann2015,Rameau2017,Follette2017,Mendigutia2018}. The ambiguity makes it difficult to place observational constraints on planet formation.

PDS 70 is currently the best system for direct studies of the planet formation process. \citet{Hashimoto2012,Hashimoto2015}  identified its complex circumstellar disk as a transitional disk with a wide gap that could be carved by planets, and \citet{Keppler2018} reported the detection of PDS 70 b within the cavity of the disk. As it was clearly inside the gap in the disk, PDS 70 b is unambiguously a planet and not a disk feature. With a stellar age estimated at $5.4 \pm 1.0$~Myr, it is one of the youngest directly imaged planets \citep{Muller2018}. It was observed to likely have H$\alpha$ emission, indicating that it was still accreting, but nearing the end of its formation process \citep{Wagner2018,Haffert2019}. Subsequently, PDS 70 c was discovered through its H$\alpha$ emission to be a second accreting protoplanet in the system, making this one of the few directly imaged multiple planet systems \citep{Haffert2019}. Follow up observations of both planets revealed mostly featureless emission spectra within current measurement uncertainties \citep{Muller2018, Mesa2019}. \citet{Muller2018} reports a possible water absorption feature between $J$- and $H$-band in PDS 70 b, although they note that it is tenuous. \citet{Christiaens2019APJ} found that the PDS 70 b spectrum has excess emission beyond 2~$\mu$m and proposed that it was surrounded by a circumplanetary disk. In ALMA mm data, \citet{Isella2019} found compact dust emission at the location of PDS 70 c suggesting it too has a circumplanetary disk. We note that \citet{Isella2019} also found another compact dust emission near the location of PDS 70 b, but significantly offset from the the planet's position. 

For both PDS 70 b and c, the constraints on their emission beyond $K$-band are weak, with the \Lp{}-band photometry of PDS 70 b reported in \citet{Muller2018} having $\approx33$\% uncertainties and the \Lp-band photometry of PDS 70 c reported by \citet{Haffert2019} possibly contaminated by circumstellar disk emission. More precise measurements at longer wavelengths are necessary to constrain the shape of the spectral energy distribution (SED) and thus the total luminosity outputted by the planets, which can provide insight into their formation history \citep{GinzburgChiang2019}. More precise measurements beyond 2~$\mu$m can also help constrain the nature of circumplanetary material, which emits at longer wavelengths \citep{Zhu2015,Szulagyi2019}.

This paper reports on the results of \Lp{}-band imaging of the PDS 70 system with Keck/NIRC2 and the newly commissioned infrared pyramid wavefront sensor \citep{Bond2018}. In Section \ref{sec:obs}, we discuss the observations and the data reductions we performed to obtain astrometry and photometry of the two planets. In Section \ref{sec:orbit}, we perform some preliminary orbital modeling of the two-planet system. In Section \ref{sec:sed}, we fit atmospheric models to the SEDs of both planets and place constraints on their radii and luminosities. In Section \ref{sec:discussion}, we use these two bulk properties in combination with evolutionary models of accreting planets to constrain the masses and mass accretion rates of the planets and discuss implications for the photospheric emission we observe.

\section{Observations and Data Reduction} \label{sec:obs}
\subsection{Observations}

We imaged PDS 70 at \Lp-band (3.426-4.126 $\mu$m) with Keck/NIRC2 on 2019 June 8 using the vortex coronagraph \citep{Vargas2016,Serabyn2017}. The average DIMM seeing was 0\farcs48. Using the 225 GHz radiometer measurements and the conversion from \citet{Dempsey2013}, we calculated that the average precipitable water vapor was 1.7~mm.  We used the infrared pyramid wavefront sensor to control the Keck adaptive optics (AO) system as part of its science verification program \citep{Bond2018}, rather than the facility Shack-Hartmann sensor. The pyramid wavefront sensor operates at $H$-band whereas the Shack-Hartmann operates at $R$-band, so it is better suited for redder stars such as PDS 70. Early commissioning data also indicated the pyramid wavefront sensor controls lower order modes better, allowing for better sensitivity within 700 mas \citep{Bond2019}.  We used the quadrant analysis of coronagraphic images for tip-tilt sensing \citep[QACITS;][]{Huby2017} algorithm to keep the star aligned behind the mask by measuring tip/tilt residuals in the NIRC2 coronagraphic images and adjusting the tip/tilt offsets between the pyramid wavefront sensor and NIRC2 accordingly. We obtained 48 frames, each consisting of 60 co-adds of 0.5~s exposures, of the star behind the vortex coronagraph. We excluded four frames from the analysis due to poor coronagraph alignment, resulting in 44 remaining frames and a total exposure time of 1320~s.  Intermittently through the observing sequence, we moved PDS 70 off of the coronagraph to take unsaturated images of the point spread function (PSF) to update the QACITS model and for photometric calibration. We took the images in pupil tracking mode to enable angular differential imaging \citep[ADI;][]{Liu2004,Marois2006}. Due to the low elevation of PDS 70 from Keck, the observing sequence provided only 28\degr{} of field rotation.

\subsection{Basic Data Reduction}\label{sec:basicdata}
\begin{figure*}
\includegraphics[width=\textwidth]{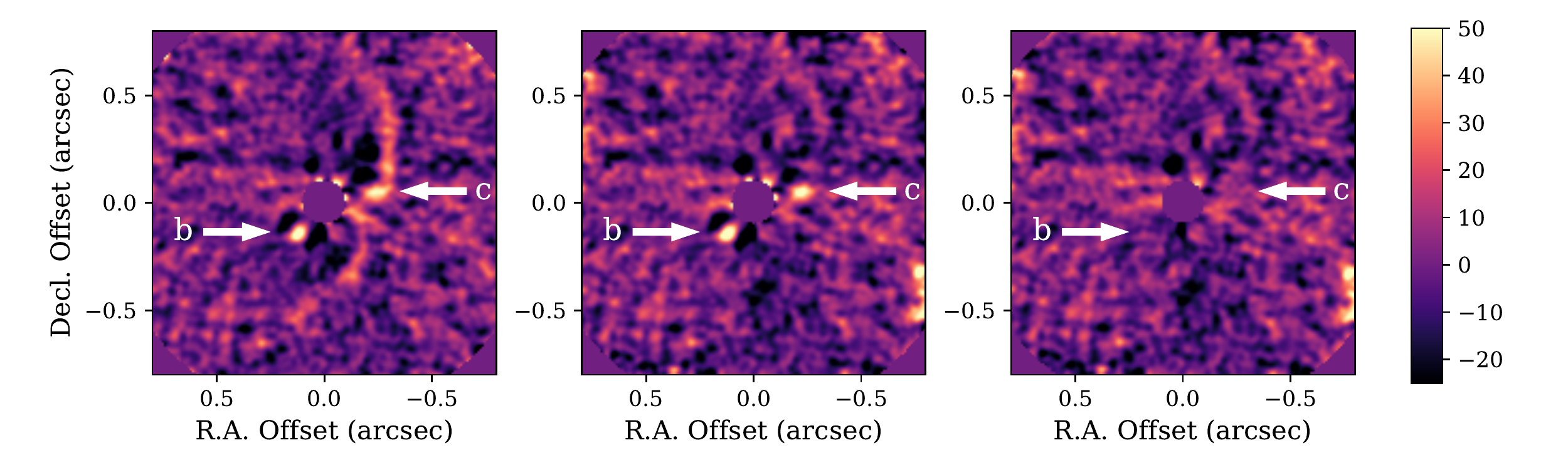}
\caption{PDS 70 in \Lp{}-band after stellar PSF subtraction. On the left is the image after regular PSF subtraction with PCA. In the middle, the image has had the disk subtracted out with a model (as described in Section \ref{sec:disksub}). On the right, the forward models for both planets (as described in Section \ref{sec:fm-bc}) have been subtracted out from the disk-subtracted image. All three images and the color bar are shown in linear scale in analog to digital units (ADU) and have been smoothed using a Gaussian kernel with a 1.5 pixel standard deviation (40\% of the width of the instrumental PSF) to average out pixel-to-pixel noise. White arrows point to PDS 70 b and PDS 70 c and are at the same location in all three images.
\label{fig:image}}
\end{figure*}

We performed initial preprocessing of the data using a general pipeline developed for NIRC2 vortex observations \citep{Xuan2018,Ruane2019}. We will briefly summarize the steps here, and we refer to reader to \citet{Xuan2018} and \citet{Ruane2019} for details. First, we corrected bad pixels and flat-field effects in each image. Then, we subtracted the thermal background from the sky and instrument using principal component analysis (PCA). Afterwards, each frame was co-registered and aligned to a common center using cross-correlation.
We then performed stellar PSF subtraction to remove the glare of the star from this preprocessed image sequence. We used the open-source Python package \texttt{pyKLIP} \citep{Wang2015} to model and subtract off the stellar glare using PCA \citep{Soummer2012}. All frames were used to construct the PCA modes, meaning each image used the same set of PCA modes for PSF subtraction. We used the first three principal components to model the star in each frame. Figure \ref{fig:image} displays the resulting image after stellar PSF subtraction. We have a clear detection of PDS 70 b. We also see the rim of the circumstellar disk with PDS 70 c right up against it.

\subsection{Disk Modeling and Subtraction}\label{sec:disksub}
Since PDS 70 c is adjacent to the circumstellar disk, we construct a model of the disk to remove it from the data in order to make unbiased measuments of PDS 70 c while minimizing contaminating flux from the disk. In our image (Figure \ref{fig:image}), we see what appears to be a partial ring, which actually is the front rim of the flared circumstellar disk seen in near-infrared scattered light images \citep{Keppler2018}. We focus on constructing a disk model to subtract out this disk component from the model, as we found that using a more complicated and physically motivated protoplanetary disk model resulted in degeneracies in the best-fit disk parameters that provided an overall worse fit to the disk we have imaged in \Lp{}-band. The disk properties have already been characterized with higher signal-to-noise data in scattered light \citep{Keppler2018} and in the mm \citep{Keppler2019}, so we instead focus on constructing a simpler model that can subtract the disk emission we see and allow us to characterize the planets. 

We construct a dust ring to model the upper rim of the disk we see in \Lp{}-band. Such a model will provide unreliable estimates of the dust spatial distribution since it only focuses on fitting this component alone, and poor constraints on dust properties since we only fit to our \Lp{}-band scattered light data. However, the inclination and position angle of the disk needs to be physical in order to reproduce the disk rim geometry.

We modeled the disk image using the radiative transfer modeling software {\tt MCFOST} \citep{Pinte2006, Pinte2009} following the technique described in \citet{Ren2019}. We caution that this analysis is designed to reproduce the observed scattering phase function, rather than the specific dust composition. To model the distribution of the light scattered by disk material, we assumed the disk is optically thin. In cylindrical coordinates, the scatterers follow a spatial distribution that is a combination of two radial power laws in the mid-plane, with a Gaussian dispersion along the perpendicular direction \citep{Augereau1999}. We assumed the scatterers are made of three compositions of dust: astronomical silicates, amorphous carbon, and H$_2$O-dominated ice \citep[][respectively]{draine84, rouleau91, li98} as in recent studies on disk modeling \citep[e.g.,][]{esposito18, Ren2019}. In radiative transfer modeling,  we calculated the distribution of scattered light using Mie theory \citep{mie1908}.
For each {\tt MCFOST} disk model, we convolved it with the NIRC2 point spread function in $L'$-band, scaled it to the NIRC2 brightness, and subtracted it from the images before PSF subtraction. We performed PCA reduction using $4$ components, then minimized the residuals in a region encompassing the disk, but excluding a circular region (10 pixel radius) where planet c resides to remove the possibility that planet c could be overfit by the model. We distributed the {\tt MCFOST} calculations using the {\tt DebrisDiskFM} package \citep{Ren2019} and used the maximum likelihood model obtained from {\tt emcee} \citep{ForemanMackey2013} as the disk model to subtract out from the images. The middle panel of Figure \ref{fig:image} shows the same stellar PSF subtraction described in Section \ref{sec:basicdata}, but done on images where the model disk was subtracted out first. We note that we find a disk inclination and position angle that are within $3^\circ$ of the values reported from mm ALMA observations \citep{Keppler2019}, which is consistent with our uncertainties on these parameters.

\begin{deluxetable*}{c|c|c}
\tablecaption{Measurements of the PDS 70 System \label{table:meas}}
\tablehead{ 
Parameter & PDS 70 b & PDS 70 c
}
\startdata 
 Epoch (MJD) & 58642 & 58642 \\ 
 Separation (mas) & $175.8 \pm 6.9$ & $223.4 \pm 8.0$ \\
 PA (\degr) & $140.9 \pm 2.2$ & $280.4 \pm 2.0$ \\
 \Lp{} Flux Ratio & $(2.05 \pm 0.34) \times 10^{-3}$ & $(9.06 \pm 3.59) \times 10^{-4}$ \\
 $\Delta$\Lp{} (mag) & $6.72 \pm 0.18$ & $7.61 \pm 0.46$ \\
 \Lp{} Flux ($10^{-17}~\rm{W}/\rm{m}^2/\mu\rm{m}$) & $7.5 \pm 1.2 $ & $3.3 \pm 1.3 $ \\ 
 \Lp{} Flux (mag) & $14.64 \pm 0.18$ & $15.5 \pm 0.46$ \\
\enddata
\end{deluxetable*}

\subsection{Forward Modeling of PDS 70 b and c}\label{sec:fm-bc}
We wish to measure the astrometry and \Lp{} photometry of PDS 70 b and c. As stellar PSF subtraction distorts the PSF of a planet, forward modeling of the signal of a planet must be done to obtain unbiased measurements. We used the KLIP-FM formalism presented by \citet{Pueyo2016} and implemented in \texttt{pyKLIP} to analytically compute the distortions on a planet PSF due to ADI and PCA. We subtract off the disk model from individual exposures to minimize any biases in the astrometry or photometry due to disk emission. We forward model each planet separately, as the planets are far enough away from each other that their signals will not distort each other.

Using the same parameters as Section \ref{sec:basicdata} to subtract off the stellar PSF, we forward modeled the distortions on PDS 70 b using an instrumental PSF from images of the star when it was moved off of the coronagraph (full width at half maximum of 8.4~pixels). We chose a 21 pixel square region centered about the approximate location of PDS 70 b and fit the forward model to the data using \texttt{emcee} \citep{ForemanMackey2013}. Measurement uncertainties were computed by creating datacubes where the signal of PDS 70 b was removed by injecting a negative planet at its location, injecting simulated planets at the same separation as PDS 70 b but different position angles, measuring their fluxes and positions, and using the scatter in the measurements of the simulated planets as the measurement uncertainties. None of these simulated planets were injected within 20 degrees of the measured location of PDS 70 b, even though we had removed it from the data, to avoid biasing the photometry of the simulated planets. Due to the close angular separation of PDS 70 b, we accounted for the transmission of the vortex coronagraph at each pixel in our forward modeled PSF. In quadrature to the error in the planet position on the detector, we also added a 4.5~mas star centering uncertainty from QACITS \citep{Huby2017}, a $0.2^\circ$ North angle uncertainty, and a 0.004~mas/pixel plate scale uncertainty \citep{Service2016}. Following \citet{Keppler2018} and \citet{Christiaens2019APJ}, we interpolated the flux of PDS 70 to the NIRC2 \Lp{}-filter (central wavelength 3.776~$\mu$m) using WISE photometry \citep{Cutri2013}, finding a star magnitude of $7.927 \pm 0.021$ and thus a planet magnitude of $14.64 \pm 0.18$.
We list our astrometric and photometric measurements for PDS 70 b in Table \ref{table:meas}. Our measured \Lp{}-band photometry for PDS 70 b is consistent with the values reported in \citet{Muller2018}, but with an error bar that is 2.3x smaller. 

We performed the same forward modeling technique to measure the astrometry and photometry of PDS 70 c. Here, PDS 70 c is adjacent to the disk signal, so subtracting off the disk signal is important for unbiased measurements. We again injected and retrieved simulated planets to estimate the uncertainties on our measurements of PDS 70 c. Using the same photometric and astrometric calibration numbers, we list our measured astrometry and photometry of PDS 70 c in Table \ref{table:meas}. We find a fainter \Lp{}-band flux ratio than \citet{Haffert2019} by 1~mag. This is likely due to the fact we removed the disk emission near the location of the planet, as the photometry for PDS 70 b agrees well between the two bodies of work, so it is unlikely a photometric calibration offset. 

We investigated potential biases introduced by the disk subtraction process. These errors would translate to additional uncertainty in PDS 70 c astrometry and photometry. In particular, we masked out the disk at the location of PDS 70 c to not overfit the planet, but this also could impact the disk model's accuracy at this location. We note that we expect this effect to be small since the scattering phase function is smooth, and the information on the disk brightness is constrained by neighboring unmasked pixels. We injected a planet in a similar location as PDS 70 c, masked a circular region around it, and repeated the disk fitting to obtain a second disk model. We then subtracted this new disk model and measured the simulated planet in the same way. We found the astrometry and photometry biases were less than the reported 1$\sigma$ uncertainties for PDS 70 c and thus consistent with the residual noise in the data. We conclude that disk fitting errors should not significantly bias our measurements.

In the right subplot of Figure \ref{fig:image}, we show the residuals of the data after subtracting off the forward model for both PDS 70 b and PDS 70 c from the image that already has the model disk removed. We do not see any systematic residuals after subtracting off the forward models.

\subsection{Extinction}\label{sec:extinction}
Given that PDS 70 resides in the Sco-Cen association \citep{Pecaut2016}, interstellar, circumstellar, and circumplanetary extinction should be considered. Following \citet{Muller2018}, we fit the visual \citep{Henden:vg,Gaia2018} and near-infrared \citep{Skrutskie:2006hla} photometry of the star to a joint set of stellar evolutionary \citep{Choi:2016kf} and atmospheric \citep{Allard2012} models. We excluded the $K$-band photometry due to an apparent 10\% excess flux at this wavelength, most likely caused by emission from circumstellar material. The fitting procedure is described in detail in \citet{Nielsen:2017kf}, although here we only fit for one star in the system. We imposed a prior on the effective temperature of the star based on the spectroscopically-derived value of $3972\pm36$\,K \citep{Pecaut2016}. We find a 3$\sigma$ upper limit on $A_V$ of 0.15\,mag, consistent with previous photometric estimates \citep{Pecaut2016,Muller2018}. This corresponds to an upper limit of 0.04 mag in $J$-band and 0.008 mag in \Lp{}-band \citep{Mathis1990}. Overall, we find that interstellar extinction should be negligible and well within measurement uncertainties of our infrared data.

For circumstellar extinction, the near-infrared scattered light data \citep{Keppler2018} and high-resolution ALMA data \citep{Keppler2019} indicate that PDS 70 b resides in a clearing in the transitional disk, so circumstellar extinction for PDS 70 b should be negligible. For PDS 70 c, it appears to be near the front rim of the circumstellar disk in projection. As PDS 70 c appears to be a point source (we are able to forward model it as a point source in Section \ref{sec:fm-bc} and the residuals look clean in Figure \ref{fig:image}), we will assume the finite size of PDS 70 c is negligible. Based on our measured astrometry, PDS 70 c lies $\sim$10~mas away from the edge of the disk when comparing to the ring model of the disk we subtracted out. This is slightly larger than our 1$\sigma$ astrometric uncertainties, so we cannot fully exclude some amount of circumstellar extinction, but the likelihood is small and the magnitude would be significantly reduced at \Lp{}-band compared to visible wavelengths. Further, \citet{Mesa2019} found that flux biases due to circumstellar dust contamination, which is directly related to extinction, to be negligble in the near-infrared at the location of PDS 70 c given the current measurement precision. In this work, the SED of PDS 70 c remains poorly constrained (see Section \ref{sec:pds70c_sed}) so if there are some small extinction effects, we would not be able to discern it. Thus, we will ignore circumstellar extinction in this work. 

For circumplanetary extinction, models can predict orders of magnitude of extinction due to circumplanetary material obscuring the disk \citep{Szulagyi2019}. The circumplanetary disks are within the Hill radii of each planet (both have $R_H \sim 2$~au using the values for semi-major axis and mass presented below in the following sections of the paper) which themselves are well below the instrumental angular resolution of any published photometry or spectrum \citep{Muller2018,Haffert2019,Christiaens2019MNRAS,Mesa2019}. Thus, we do not try to measure circumplanetary extinction, but rather aim to characterize the total emission coming from the planet and any circumplanetary material. When comparing our measured luminosities to the evolutionary models from \citet{GinzburgChiang2019} in Section \ref{sec:evo_models}, what we use is the total luminosity from both components combined, so this approach is fully consistent with the model assumptions.

\section{Orbital Constraints}\label{sec:orbit}
With the single additional astrometric epoch, the orbit remains relatively unconstrained. We expect a large degenerate set of orbits. Many of these are unlikely to be physical if the orbits of planets b and c cross, or if they are too misaligned from one another. There is also no noticeable warp in the disk, so we expect the planets to be approximately coplanar with the circumstellar disk. Because of this, we do not simply fit two Keplerian orbits to the data, since most of the orbits will likely not reflect reality. Instead, we impose physically motivated priors to constrain the fit. 

We use the same orbital parameter set as \citet{Wang2018}, but the reference epoch for $\tau$ is MJD 58,849 (2020 January 1st). Orbital parameters corresponding to PDS 70 b and c are denoted by their respective subscripts. We start out with uninformative priors on most of the orbital parameters, which are listed in Table \ref{table:orbitparams}. We used a Gaussian prior for parallax based on the parallax of $8.8159 \pm 0.0405$~mas from \textit{Gaia} DR2 \citep{Gaia2018}. We used a Gaussian prior for the total mass of the system of $0.760 \pm 0.078$~$M_\odot$ based on the mass derived by \citet{Muller2018}, but with an additional 10\% uncertainty to account for potential systematics in the photometrically derived mass.

We then added additional priors that constrain the stability of the system. We require that orbits cannot cross, so that the periastron of PDS 70 c is always larger than the apastron of PDS 70 b:

\begin{equation}
    a_c (1 - e_c) > a_b (1 + e_b).
\end{equation}
We give uniform weight to orbits that satisfy this criterion and reject orbits that do not. \citet{Haffert2019} hypothesized the planets, assuming they were coplanar, could be packed closely enough to be in or near the 2:1 mean-motion resonance. For massive gas giants at these large separations, \citet{Wang2018} found that stable orbits of the HR 8799 planets, which also are in or near 2:1 mean-motion resonances, required their orbital planes to be within $8^\circ$ of coplanar. However, that work did not fully explore parameter space so there might be some stable orbits that are more inclined. We define mutual inclination, $\Phi_{12}$, between orbital plane 1 and 2 with the same notation as \citet{Bean2009}:

\begin{equation}
    \cos (\Phi_{12}) = \cos(i_1)\cos(i_2) + \sin(i_1)\sin(i_2)\cos(\Omega_1 - \Omega_2).
\end{equation}
Here, $i$ and $\Omega$ describe the inclination and the position angle of the ascending node for each plane. We add a prior that prefers orbital configurations in which the orbital planes of PDS 70 b, PDS 70 c, and the circumstellar disk are more coplanar. We place more conservative constraints on coplanarity than the upper limit of $8^\circ$ found by \citet{Wang2018}. For each pair of orbital planes, we apply a Gaussian prior on $\Phi$ centered at $0^\circ$ with a standard deviation of $10^\circ$. For the orbital plane of the disk, we fix the inclination to $128.3^\circ$, which is the same $51.7^\circ$ reported in \citet{Keppler2019} but for clockwise orbits, and the position angle of the ascending node to $156.7^\circ$. We note that the velocity maps of the gas in the circumstellar disk break the $180^\circ$ degeneracy in $\Omega$. Since we have three orbital planes, this results in three Gaussian priors, one for each mutual inclination between two of the planes, to constrain four orbital parameters ($i_b$, $\Omega_b$, $i_c$, $\Omega_c$).

For orbit fitting, we use an unreleased version of \texttt{orbitize!} \citep{Blunt2020} with commit hash \texttt{361764} that supports fitting multiple planets. In addition to our measured NIRC2 point, we use the published PDS 70 b astrometry from \citet{Muller2018}, the published PDS 70 c astrometry from \citet{Mesa2019}, and the H$\alpha$ astrometry of both planets from \citet{Haffert2019}. We use the parallel-tempered affine-invariant sampler implemented in \texttt{ptemcee} \citep{ForemanMackey2013,Vousden2016} with 20 temperatures and 1000 walkers per temperature. Each walker discarded the first 5000 steps as a ``burn-in" phase, and obtained 500 samples of the posterior after only saving every tenth step to minimize correlation between consecutive samples. This resulted in 500,000 samples of the posterior. Convergence of the walkers was determined by requiring the burn-in phase to be more than 10 autocorrelation times and through visual inspection of the chains as discussed in \citep{Blunt2020}.

\begin{deluxetable}{c|c|c|c}
\tablecaption{Orbital Parameters for PDS 70 b and c \label{table:orbitparams}}
\tablehead{ 
Orbital Element  & Prior & 95\% CI & Best Fit 
}
\startdata 
 $a_b$ (au)               & LogUniform(1, 100)\tablenotemark{a} & $20_{-4}^{+3}$ & 24 \\
 $e_b$                    & Uniform(0, 1)\tablenotemark{a} & $0.19^{+0.30}_{-0.18}$ & 0.17 \\
 $i_b$ (\degr)            & $\sin(i)$\tablenotemark{b} & $140^{+13}_{-12}$ & 138 \\
 $\omega_b$ (\degr)       & Uniform(0, 2$\pi$) & $148 \pm 62$ & 84 \\
 $\Omega_b$ (\degr)       & Uniform(0, 2$\pi$)\tablenotemark{b} & $159^{+17}_{-19}$ & 162 \\
 $\tau_b$                 & Uniform(0, 1) & $0.30^{+0.20}_{-0.15}$ & 0.12 \\
 $a_c$ (au)               & LogUniform(1, 100)\tablenotemark{a} & $34_{-6}^{+12}$ & 40 \\
 $e_c$                    & Uniform(0, 1)\tablenotemark{a} & $0.11^{+0.24}_{-0.11}$ & 0.09 \\
 $i_c$ (\degr)            & $\sin(i)$\tablenotemark{b} & $132_{-13}^{+14}$ & 130 \\
 $\omega_c$ (\degr)       & Uniform(0, 2$\pi$) & $136_{-115}^{+100}$ & 218 \\
 $\Omega_c$ (\degr)       & Uniform(0, 2$\pi$)\tablenotemark{b} & $156_{-22}^{+23}$ & 162 \\
 $\tau_c$                 & Uniform(0, 1) & $0.74^{+0.24}_{-0.38}$ & 0.92 \\
 Parallax (mas)         & $\mathcal{N}$(8.8159, 0.0405) & $8.819 \pm 0.08$ & 8.818 \\
 $M_{tot}$ ($M_\odot$)  & $\mathcal{N}$(0.76, 0.079) & $0.79 \pm 0.15$ & 0.78 \\
\enddata
\tablecomments{The 95\% credible interval values (95\% CI) are centered about the median, and the subscript and superscript denote the range spanned by the 2.5 and 97.5 percentile values. The best fit column lists the fit with the maximum posterior probability. We note that the best fit orbit is generally not a good estimate of the true orbit, but can be useful as a representative orbit whereas the median of all the values is not always a valid orbit due to strong correlations in the orbital parameters. }
\tablenotetext{a}{Additional prior on periastron of c is larger than apastron of b}
\tablenotetext{b}{Additional Gaussian prior on the coplanarity of b, c, and the disk}
\end{deluxetable}

\begin{figure*}
\plotone{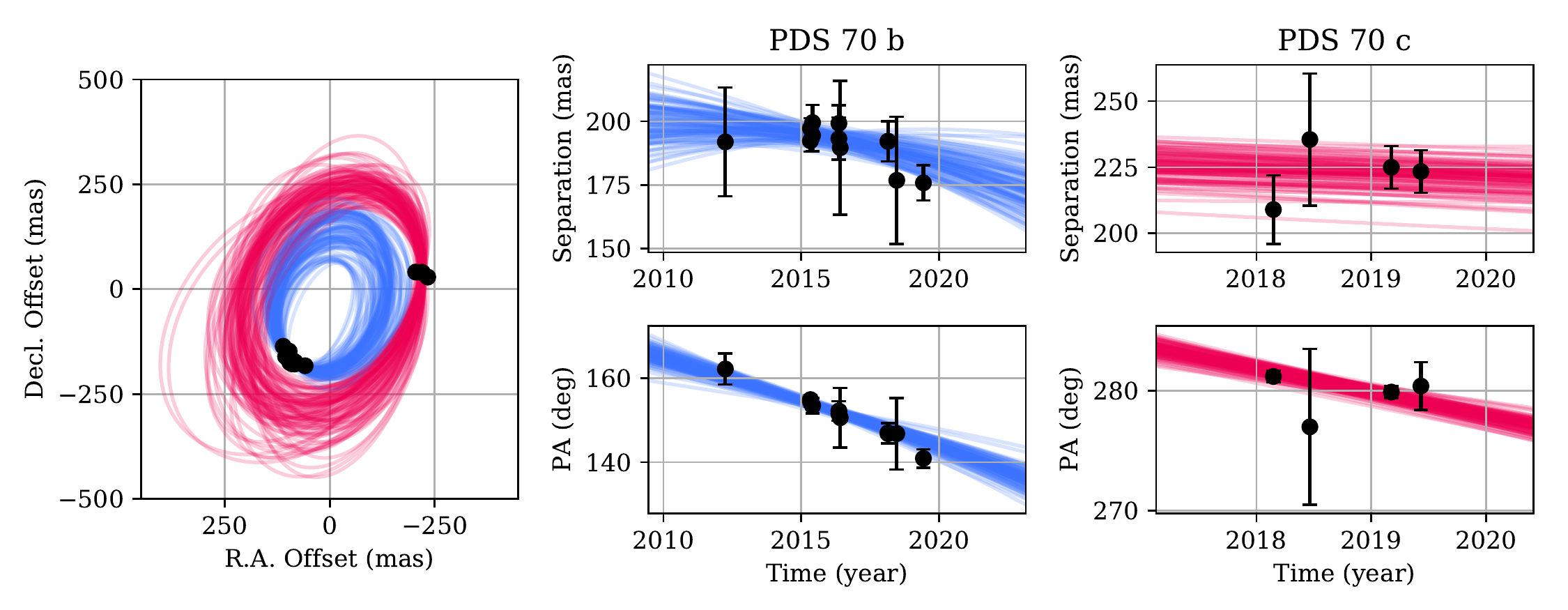}
\caption{The orbits of PDS 70 b and c. On the left, 100 randomly drawn orbits from the posterior are plotted in their sky-projection with blue lines corresponding to PDS 70 b and red lines corresponding to PDS 70 c. Measured astrometry are plotted in black. On the right four plots, the same randomly drawn orbits are plotted as a function of time for both planets, and the measured astrometry used in the fit are plotted with measurement errors. The Keck point reported in this work is the point most recent in time in the plots. 
\label{fig:orbit}}
\end{figure*}

We plot the orbit fit in Figure \ref{fig:orbit} and list the 95\% credible range of each orbital parameter in Table \ref{table:orbitparams}. Due to the strong covariances in the parameters, we also list the best fit orbit simply as a valid representative orbit for reference. We note that the best fit orbit in situations like this is generally not a good estimate of the true orbit due to overfitting a short orbital arc with 6 orbital elements, but can be useful for near-term orbit prediction. 

We find period ratios between PDS 70 c and PDS 70 b to be in the 95\% credible interval of 1.5 to 3.9. The planets could be in mean-motion resonance as hypothesized by \citet{Haffert2019}. Due to the coplanarity constraint we placed on the orbital planes of the two planets and the fact the orbital planes were nearly unconstrained by current astrometry, we find that the mutual inclinations of each pair of orbital planes between PDS 70 b, PDS 70 c, and the disk all have a 95\% credible interval from $2^\circ$ to $23^\circ$ that is dominated by our prior. Rigorous stability constraints would help reduce the parameter space of possible orbits \citep{Wang2018}. We defer such analysis to future work with more astrometric measurements to constrain the orbit and reduce the parameter space of possible orbits to search.

\section{SED Fitting}\label{sec:sed}
To study the atmosphere and accretion history of PDS 70 b and c, we analyze the SED of the planets to infer luminosities and radii and compare them to the accreting planet evolutionary models presented in \citet{GinzburgChiang2019}. Given that these two planets are unlike other directly imaged planets and brown dwarfs in that they appear to still be accreting from the circumstellar disk \citep{Wagner2018, Haffert2019}, we note that it is very likely that no existing atmospheric model accurately describes its SED. With that in mind, the main focus of this work is to measure the luminosities and radii of the two planets, and acknowledge that there are likely errors and biases in the inferred quantities beyond the formal errors from the fits. We aim to mitigate this by averaging over all models that are equally adequate fits to the data and by noting that the evolutionary models are not extremely sensitive to the exact values (see Section \ref{sec:evo_models}). 

\subsection{PDS 70 b SED}\label{sec:pds70b_sed}
In addition to the \Lp{} photometry reported in this work, we include the $R\sim30$ $YJH$ SPHERE spectrum and $K$- and \Lp{}-band photometry reported in \citet{Muller2018} and the $R\sim100$ $K$-band SINFONI spectrum from \citet{Christiaens2019MNRAS}. With $\sim$2x smaller uncertainties on the \Lp{} photometry than \citet{Muller2018}, we expect better constraints on the temperature, radius, and luminosity of the planet, as this longer wavelength point helps constrain the overall spectral shape of the planet's SED. We fit multiple models to the data to explore different assumptions and to quantify model biases.

First, we fit a simple blackbody to the SED. Given that the only evidence of molecular absorption is a tentative water absorption feature between $J$- and $H$-band measured by SPHERE \citep{Muller2018}, a simple model like a blackbody could be a good fit to the data, possibly resulting from  an accreting dust shell shrouding the planet. We model the flux received, $F_\lambda$, by the equation

\begin{equation}
    F_\lambda = \frac{\pi R_{b}^2}{d^2} B_\lambda(T_{b})
\end{equation}

\noindent where $R_{b}$ is the radius of PDS 70 b, $T_{b}$ is the temperature of the blackbody, $d$ is the distance to the planet, and $B_\lambda$ is the specific intensity of a blackbody. Note that in the following section for PDS 70 c, we will use $R_c$ and $T_c$ to refer to its respective radius and temperature.

We adopt a Gaussian likelihood function to fit the model to the data. For both the SPHERE and SINFONI spectra, the noise is likely correlated between nearby spectral channels given that the scatter between adjacent spectral channels is smaller than the reported uncertainties. This is not surprising since correlated noise due to spectral oversampling and speckle noise has been reported in high-contrast observations with many integral field units \citep{DeRosa2016,Samland2017,Currie2018}. Thus, we assume the total reported uncertainty is a combination of correlated and uncorrelated noise added in quadrature. We adapt the framework from \citet{Czekala2015} for fitting stellar spectra in the presence of correlated noise to fitting the spectra of these planets. We model the correlated noise for each spectrum as a separate Gaussian process parameterized by a square exponential kernel:

\begin{equation}
\begin{split}
    C_{ij} = & (f_{amp} \sigma_{i}) (f_{amp} \sigma_{j}) \exp\left(\frac{-(\lambda_{i} - \lambda_{j})^2}{2 l^2}\right)  \\
    & + (1-f_{amp}^2) \sigma_i^2\delta_{ij}.
\end{split}
\end{equation}
Here $C_{ij}$ is the element of the covariance matrix corresponding to wavelength channels $i$ and $j$, $\sigma_{i}$ is the measured uncertainty in channel $i$, $\lambda_i$ is the wavelength of that channel, $l$ is the correlation length, $f_{amp}$ is the fraction of the measured uncertainty that is due to correlated noise, and $\delta_{ij}$ is the Kronecker delta. Given that the total error is measured, we need to find the fractional error that is due to correlated noise to set the amplitude of the correlated noise. The rest is uncorrelated noise that only appears in the diagonal of the covariance matrix. We note this treatment of the Gaussian process amplitude differs from \citet{Czekala2015} as their reported errors correspond only to the uncorrelated noise term whereas ours encompass both. For each dataset, we fit for $f_{amp}$ and $l$ in order to characterize the correlated noise. Otherwise, treating correlated noise as uncorrelated noise will bias the posteriors, such as making them more constrained than in reality, unjustly over-weighing them over single photometric points, or favoring spurious spectral features in the models \citep{Greco2016}. 

\begin{figure*}
    \centering
    \includegraphics{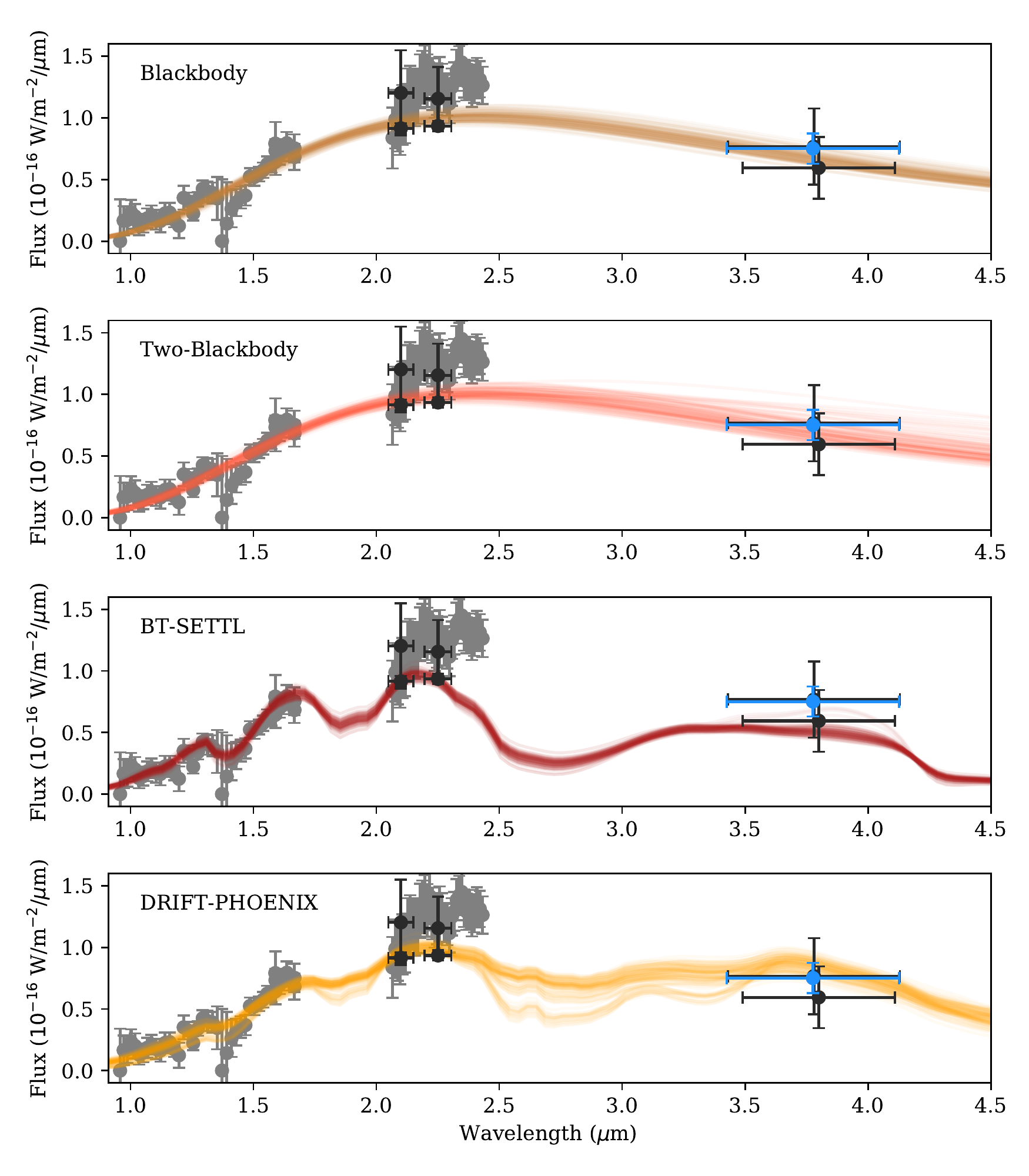}
    \caption{Fits to the spectral energy distribution of PDS 70 b. From top to bottom, each of the four panels shows 100 possible fits (chosen at random from their respective posteriors) for blackbody (brown), two-blackbody (peach), BT-SETTL (maroon), and DRIFT-PHOENIX (yellow) models respectively. In all four panels, the blue point is the Keck \Lp{} photometry measured in this work, the black points are literature photometry used in the fit, and the gray points are literature spectra used in the fit. The error bars in the y-axis denote 1$\sigma$ errors while the horizontal bars indicate the bandpass of the photometric points.}
    \label{fig:atm_fits}
\end{figure*}

We performed Bayesian parameter estimation using the \texttt{emcee} package. In addition to the two model parameters of the blackbody model, the radius and temperature, we fit for four nuisance parameters that quantify systematics in the data: the amplitude and correlation length for the Gaussian process that describes the correlated noise in the SPHERE IFS data and the amplitude and correlation length of the correlated noise in the SINFONI data. We noticed that the SINFONI spectrum is noticeably offset from the SPHERE IRDIS $K$-band photometry, so there will be inherent disagreement in $K$-band in our fits. We used 100 walkers in our affine-invariant sampler, burned each walker in for 500 steps, and used 200 following steps from each walker to construct a posterior with 20,000 samples. Convergence was assessed through visual inspection of the chains. The fit to a single blackbody are plotted in the top panel of Figure \ref{fig:atm_fits}. The 95\% credible intervals for the parameters are listed in Table \ref{table:atm_models}. We note that we report 95\% credible intervals rather than the standard 68\% ranges to express the full range of uncertainties in model parameters rather than formal ``1$\sigma$" uncertainties since there are likely model biases. The posterior for the planet's luminosity, $L_b$, was derived by computing the blackbody luminosity for each set of model parameters in our sampled posterior using the equation
\begin{equation}
    L_b = 4\pi R_b^2 \sigma_{SB} T_b^4
\end{equation}
where $\sigma_{SB}$ is the Stefan-Boltzmann constant. We also list the median value and 95\% credible interval for luminosity posterior in Table \ref{table:atm_models}. 

We also explore a two-blackbody model, which emits flux 

\begin{equation}
    F_\lambda = \frac{1}{d^2} \left( \pi R_{b}^2 B_\lambda(T_{b}) +  \pi R_{2}^2 B_\lambda(T_{2}) \right)
\end{equation}
where $R_2$ and $T_2$ is the radius and temperature of the second blackbody component. The two additional model parameters bring the number of free parameters to eight. This second blackbody could trace circumplanetary material, as hypothesized by \citet{Christiaens2019MNRAS}. In this work, we are agnostic to the exact nature of this second component, and merely explore whether including it can lead to better fits to the data. The second blackbody could also improve derived values from the \citet{GinzburgChiang2019} accreting planet model, which is based on energy balance; the second blackbody will simulate energy from accretion reprocessed and radiated away at longer wavelengths that is not accounted for in a single blackbody model fit (e.g., due to circumplanetary dust).

We also fit the SED to two grids of atmospheric models: the BT-SETTL atmospheric model grid \citep{Allard2012} and the DRIFT-PHOENIX model grid \citep{Woitke2003, Woitke2004, Helling2006, Helling2008}. In addition to the six parameters fit in the single blackbody fit, we also vary the surface gravity ($\log_{10}(g)$ in cgs units) for both of these model grids. Note that the temperature parameter of these two atmospheric models correspond to the effective temperature of the model SED. For the DRIFT-PHOENIX models, we also vary metallicity ($[M/H]$) since the grid of models provides a limited range in $[M/H]$. We included these parameters in our fit, using uniform priors with bounds dictated by the limits of the grids. BT-SETTL has a range of surface gravities from 3.5 to 5.5 (steps of 0.5 in the grid). DRIFT-PHOENIX has a range of surface gravities between 3.0 and 5.5 (steps of 0.5), and a range of metallicities between $-0.3$ and $0.3$ (steps of 0.3). For both grids, due to the 1000 K lower bound, we considered a range of effective temperatures between 1000 and 1500 K (steps of 100 K in both grids). To generate spectra between grid points, we used linear interpolation of the closest grid models. We note that such model atmospheres have struggled to match the broad-band SEDs of field brown dwarfs with temperatures similar to the PDS 70 planets \citep[e.g.,][]{Marocco2014,liu2016}, likely due to challenges of modeling condensate clouds, and we might expect similar difficulties to be seen in our analysis here.

We performed the Bayesian parameter estimation for these three models with the affine-invariant sampler in \texttt{emcee}. We used 100 walkers and obtained 600 samples from each walker after discarding the first 900 samples as an initial burn in. The 95\% credible intervals about the median are listed in Table \ref{table:atm_models}. The two-blackbody, BT-SETTL, and DRIFT-PHOENIX models are plotted in Figure \ref{fig:atm_fits}. We also derived the luminosity posteriors for each model based on our posterior of sampled parameters. For the two-blackbody model, the luminosity was calculated as a sum of single blackbody luminosities:
\begin{equation}
    L_b = 4\pi R_b^2 \sigma_{SB} T_b^2 + 4\pi R_2^2 \sigma_{SB} T_2^2.
\end{equation}
For the model grids, there is no analytical equation. For each set of parameters from our sampled posterior, we compute the corresponding bolometric luminosity by numerically integrating the model spectrum $F_{\lambda_i,model}$ over the entire wavelength range provided by the model at its native spectral resolution with wavelength spacing per spectral channel $\delta \lambda_i$ and multiplied it by the surface area:
\begin{equation}
    L_b = 4 \pi R_b^2 \sum_i F_{\lambda_i,model} \delta \lambda_i.
\end{equation}
The native spectral resolution of models is high (R $> 10,000$), so the numerical errors due to this integration are negligible. 
We list the median and 95\% credible intervals of the derived luminosity posteriors for each model in Table \ref{table:atm_models}. Due to a combination of the limited range in surface gravities and metallicities of these models and weak constraints on these parameters due to the quality of existing data, the data are consistent with surface gravities and metallicities across the entire parameter range. For the BT-SETTL model, we see a preference towards having a surface gravity at the lower bound of the model grid. For this work, we will marginalize our fits across these parameters and focus on the effective temperature and radius of each model.

We use the Akaike information criterion (AIC) to determine the relative goodness of fit of these models \citep{Akaike1973,Burnham}. For each model, we consider the parameters of that model that give the lowest AIC (i.e., the maximum likelihood model). We consider the single blackbody model as the fiducial model, as it is the simplest model we considered. We compute the difference between the other models and the blackbody fit by $\Delta \textrm{AIC} = \textrm{AIC}_{model} - \textrm{AIC}_{blackbody}$. We list these values in Table \ref{table:atm_models}. We find that the single blackbody model is the preferred model based on the AIC. The two-blackbody model has slightly less support from the data, as the additional two parameters do not significantly improve the fit. The BT-SETTL model does not fit the new \Lp{} photometry. It has a $\Delta \textrm{AIC} > 10$, which implies there is no support for this model compared to the other models considered \citep{Burnham}. We find that the DRIFT-PHOENIX model has considerably less empirical support for it compared to the blackbody models, but remains under the threshold for exclusion ($\Delta \textrm{AIC} < 10$). We note that this analysis does not imply that a single blackbody is the correct model. Rather, the more sophisticated models explored in this work do not do a better job given the number of additional free parameters they introduce. It is very likely that a single blackbody is not the true SED of this planet, but additional data is necessary to justify using more complex models.

Focusing on the three models (blackbody, two-blackbody, DRIFT-PHOENIX) that fit the data the best, we find that there is some disagreement in the derived radius and temperature. The single blackbody model prefers lower temperatures but larger radii, while DRIFT-PHOENIX prefers the opposite, and the two-blackbody model is somewhere in between. However, all of the models place the radius of the photosphere between 2-3~$R_{Jup}$. This is significantly larger than the 1.5 to 1.8 $~R_{\textrm{Jup}}$ predicted by hot-start evolution models of isolated planets between 1 and 10 $M_{\textrm{Jup}}$ \citep{Baraffe2003}, and could be due to possible emission from lower pressure levels from accreting material shrouding the planet. Indeed, the median $T_2$ and $R_2$ values of the two-blackbody model are $\sim$700~K and $\sim$5~$R_{Jup}$, respectively, which may be from circumplanetary material. However, we note that the large uncertainties on this second component indicate that this is a tentative interpretation that relies heavily on the single \Lp{} photometric point reported in this paper. Alternatively, the large radius could be the result of high atmospheric opacity slowing down the planet's contraction (\citealt{GinzburgChiang2019} and Section \ref{sec:discussion}). 

The uncertainties of the derived luminosities of the three models all overlap. In fact, the single blackbody and DRIFT-PHOENIX models have total luminosities that agree to within 10\%. The large positive tail in the luminosity inferred using the two-blackbody model is due to the second component being relatively unconstrained. Our tight constraint on the total luminosity is due to having adequate sampling of the SED over the 1-4~$\mu$m spectral region, which covers the bulk of the emission from the planet. If we average the luminosity posteriors of the blackbody, two-blackbody, and DRIFT-PHOENIX models assuming equal weight, we find a model-averaged luminosity posterior of  $1.48^{+0.58}_{-0.30} \times 10^{-4} L_\odot$ (95\% credible interval). We will use this luminosity in Section \ref{sec:evo_models} to infer a mass and mass accretion rate.

Even though the BT-SETTL model was a relatively poor fit to the data, we can directly compare the parameters we estimated to those for the same model from \citet{Muller2018} and \citet{Christiaens2019APJ}. We find that our derived effective temperature is lower by 200-400~K, while our derived radius is in between those two previous works. If we compare the better fitting DRIFT-PHOENIX and blackbody models to the suite of model fits in \citet{Muller2018}, we find good agreement in the radius, but we prefer effective temperatures that are higher by 100-200~K. While we do not fit any circumplanetary disk models to the data other than a simple two-component blackbody in this work, our \Lp{} flux is consistent with the predicted flux from the circumplanetary disk model in \citet{Christiaens2019APJ}. However, we do not find that the quality of the current data requires including this additional component in the SED. 

\begin{deluxetable*}{c|c|c|c|c|c}
\tablecaption{Model fits to SED of PDS 70 b\label{table:atm_models}}
\tablehead{ 
 Parameter & Prior & Blackbody & Two-Blackbody & BT-SETTL & DRIFT-PHOENIX
}
\startdata 
$T_{b}$ (K) & \shortstack[c]{Uniform(100, 2500)\tablenotemark{a} or\\ Uniform(1000, 1500)\tablenotemark{b,c}}  & $1204^{+52}_{-53}$ & $1218^{+112}_{-64}$ & $1243^{+31}_{-63}$ & $1346^{+75}_{-136}$ \\
$R_b$ ($R_{\textrm{Jup}}$) & Uniform(0.5, 5) & $2.72^{+0.39}_{-0.34}$ & $2.62^{+0.48}_{-0.81}$ & $1.93^{+0.26}_{-0.08}$ & $2.09^{+0.23}_{-0.31}$ \\
$T_2$ (K) & Uniform(100, 2500) & - &  $520^{+533}_{-396}$ & - & - \\
$R_2$ ($R_{\textrm{Jup}}$) & Uniform(0.5, 10) & - & $4.49^{+5.16}_{-3.79}$ & - & - \\
$\log(g)$ (cgs) & Uniform(3.0\tablenotemark{b}/3.5\tablenotemark{c}, 5.5) & - & - & $3.51^{+0.08}_{-0.01}$\tablenotemark{d} & $4.01^{+1.17}_{-0.96}$\tablenotemark{d} \\
$[M/H]$ & Uniform(-0.3, 0.3) & - & - & - & $-0.01^{+0.29}_{-0.27}$\tablenotemark{d} \\
$L_b$ ($10^{-4}~L_\odot$) & Derived & $1.48^{+0.16}_{-0.15}$ &   $1.59^{+0.63}_{-0.21}$ & $0.86^{+0.6}_{-0.5}$ & $1.36^{+0.19}_{-0.27}$ \\
\hline
SPHERE IFS $f_{amp}$ & LogUniform($10^{-5}$, 1) & $0.81^{+0.13}_{-0.81}$ & $0.80^{+0.14}_{0.80}$ & $0.66^{+0.22}_{-0.66}$ & $0.78^{+0.17}_{-0.69}$ \\
SPHERE IFS $l$ ($\mu$m) & LogUniform($10^{-3}$, 0.5) & $0.059^{+0.057}_{-0.050}$ & $0.059^{+0.018}_{-0.054}$ & $0.13^{+0.33}_{-0.12}$ & $0.062^{+0.243}_{-0.048}$ \\
SINFONI $f_{amp}$ & LogUniform($10^{-5}$, 1) & $0.82^{+0.08}_{-0.82}$ & $0.83^{+0.07}_{-0.83}$ & $0.02^{+0.70}_{-0.02}$ & $0.76^{+0.12}_{-0.76}$ \\
SINFONI $l$ ($\mu$m) & LogUniform($10^{-3}$, 0.5) & $0.176^{+0.172}_{-0.110}$ & $0.171^{+0.168}_{-0.107}$ & $0.158^{+0.142}_{-0.085}$ & $0.182^{+0.163}_{-0.111}$ \\
\hline
$\Delta$AIC & Derived & 0 & 2.56 & 34.57  & 7.51 \\
\enddata
\tablecomments{For each parameter, a 95\% credible interval centered about the median is reported. The superscript and subscript denote the upper and lower bounds of that range. }
\tablenotetext{a}{Blackbody/Two-Blackbody bound}
\tablenotetext{c}{DRIFT-PHOENIX bound}
\tablenotetext{b}{BT-SETTL bound}
\tablenotetext{d}{Parameter hits bound of prior, which were imposed due to available parameter space of model grid}
\end{deluxetable*}

\subsection{PDS 70 c SED}\label{sec:pds70c_sed}
\begin{figure*}
    \centering
    \includegraphics{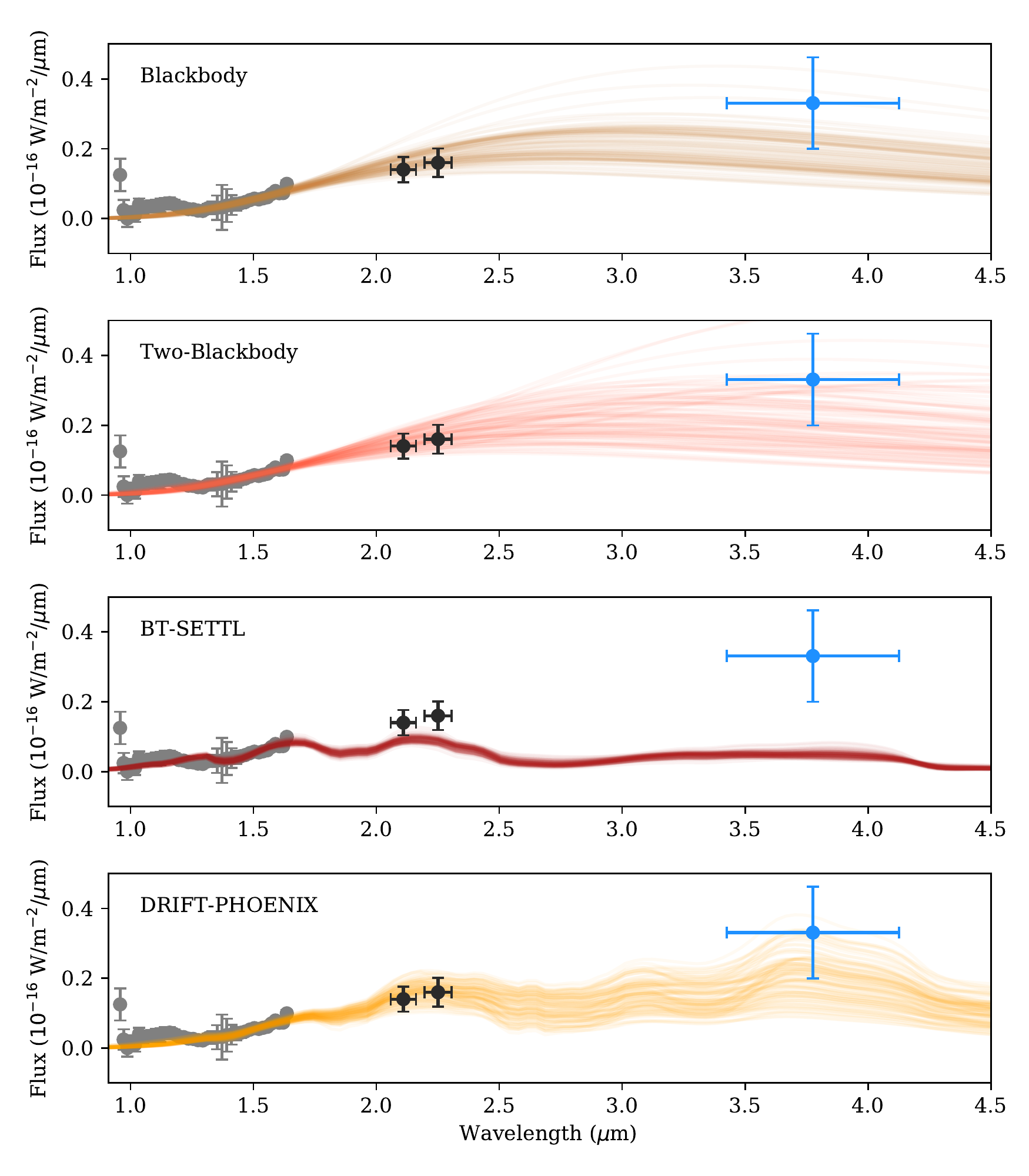}
    \caption{Same as Figure \ref{fig:atm_fits}, but for PDS 70 c.}
    \label{fig:c_atm_fits}
\end{figure*}

We repeat the same SED fitting process for PDS 70 c. We use $R\sim30$ near-infrared spectrum and $K$-band photometry measured by SPHERE that are reported in \citet{Mesa2019} in addition to our \Lp{}-band photometric point. We do not use the photometry reported by \citet{Haffert2019}, as it is unclear how much of the photometry is contaminated by disk emission. We continue to use a Gaussian process to model any correlated noise component in the SPHERE spectrum using a square exponential kernel. We fit the same four models to the measured data using the same procedure as for PDS 70 b. The model fits are plotted in Figure \ref{fig:c_atm_fits} and the 95\% credible intervals of the model parameters are listed in Table \ref{table:c_atm_models}. Note that we replaced the subscript b with subscript c to denote PDS 70 c. 

We again find that the fiducial blackbody model is the preferred model based on the AIC. The two-blackbody and DRIFT-PHOENIX models also have some support from the data, but the BT-SETTL models ($\Delta \textrm{AIC} > 10$) do not, as they underpredict the \Lp{} photometry. The three better-fitting models favor a PDS 70 c that is cooler than PDS 70 b by $\sim$200~K and more compact in radius. The model parameters are less well constrained for PDS 70 c, so it is difficult to interpret the values of individual parameters in much detail, as many are only marginally constrained. 

If we marginalize over all of the parameters and look at the total luminosity inferred from each model, we find that PDS 70 c is less luminous than PDS 70 b by a factor of $\sim$3, though we are essentially only able to constrain the order of magnitude of the luminosity from the planet. The lower luminosity of PDS 70 c, as inferred from its total integrated SED, is consistent with its similarly lower H$\alpha$ emission as compared to that of PDS 70 b \citep{Haffert2019}. If we average the luminosity posteriors of the blackbody, two-blackbody, and DRIFT-PHOENIX models assuming equal weight, we find an average luminosity of $3.60^{+5.84}_{-1.93} \times 10^{-5} L_\odot$, where the quoted range is the 95\% credible interval. We will use this average luminosity posterior in Section \ref{sec:evo_models}.

\begin{deluxetable*}{c|c|c|c|c|c}
\tablecaption{Model fits to SED of PDS 70 c\label{table:c_atm_models}}
\tablehead{ 
 Parameter & Prior & Blackbody & Two-Blackbody & BT-SETTL & DRIFT-PHOENIX
}
\startdata 
$T_{c}$ (K) &  \shortstack[c]{Uniform(100, 2500)\tablenotemark{a} or\\ Uniform(1000, 1500)\tablenotemark{b,c}} & $995^{+141}_{-97}$ & $1030^{+289}_{-216}$ & $1251^{+129}_{-104}$ & $1202^{+156}_{-160}$ \\
$R_c$ ($R_{\textrm{Jup}}$) & Uniform(0.5, 5) & $2.04^{+1.22}_{-0.89}$ & $1.65^{+1.46}_{-1.10}$ & $0.59^{+0.17}_{-0.08}$ & $1.13^{+0.56}_{-0.43}$ \\
$T_2$ (K) & Uniform(100, 2500) & - &  $544^{+521}_{-421}$ & - & - \\
$R_2$ ($R_{\textrm{Jup}}$) & Uniform(0.5, 10) & - & $4.44^{+5.23}_{-3.68}$ & - & - \\
$\log(g)$ (cgs) & Uniform(3.0\tablenotemark{b}/3.5\tablenotemark{c}, 5.5) & - & - & $3.60^{+0.47}_{-0.09}$\tablenotemark{d} & $3.75^{+1.47}_{-0.71}$\tablenotemark{d} \\
$[M/H]$ & Uniform(-0.3, 0.3) & - & - & - & $-0.00^{+0.28}_{-0.29}$\tablenotemark{d} \\
$L_c$ ($10^{-4}~L_\odot$) & Derived & $0.39^{+0.27}_{-0.17}$ &   $0.49^{+0.66}_{-0.26}$ & $0.083^{+0.015}_{-0.016}$ & $0.27^{+0.16}_{-0.12}$ \\
\hline
SPHERE IFS $f_{amp}$ & LogUniform($10^{-5}$, 1) & $0.77^{+0.11}_{-0.21}$ & $0.76^{+0.11}_{0.76}$ & $0.80^{+0.14}_{-0.80}$ & $0.76^{+0.13}_{-0.76}$ \\
SPHERE IFS $l$ ($\mu$m) & LogUniform($10^{-3}$, 0.5) & $0.111^{+0.131}_{-0.077}$ & $0.104^{+0.135}_{-0.078}$ & $0.040^{+0.069}_{-0.024}$ & $0.094^{+0.201}_{-0.077}$ \\
\hline
$\Delta$AIC & Derived & 0 & 3.48 & 14.00 & 4.36 \\
\enddata
\tablecomments{For each parameter, a 95\% credible interval centered about the median is reported. The superscript and subscript denote the upper and lower bounds of that range. }
\tablenotetext{a}{Blackbody/Two-Blackbody bound}
\tablenotetext{c}{DRIFT-PHOENIX bound}
\tablenotetext{b}{BT-SETTL bound}
\tablenotetext{d}{Parameter hits bound of prior, which were imposed due to available parameter space of model grid}
\end{deluxetable*}

\section{Discussion}\label{sec:discussion}
\subsection{Evolutionary Models}\label{sec:evo_models}

We translated the inferred luminosities of PDS 70 b and c to planet masses and accretion rates using the model of \citet{GinzburgChiang2019}, who evolved planet radii and luminosities following an initial rapid phase of runaway growth up to the eventual dispersal of the protoplanetary disk. This model postulates that, as planet accretion rates diminish, presumably as a result of gap opening, planets simultaneously contract and accrete such that their thermal cooling times remain equal to their growth times. The Kelvin--Helmholtz cooling time is calculated by modeling the planet with a radiative envelope and a convective interior, where regions of partial ionization and dissociation are resolved in order to obtain an accurate density profile. We treated the opacity $\kappa$ at the radiative--convective boundary, which dictates the cooling and contraction rate, as a free parameter to accommodate uncertainties in the physics of dust growth and sedimentation in the planet's atmosphere \citep{Movshovitz2010,Mordasini2014,Ormel2014}. Specifically, we varied the opacity from a dust free $\kappa=10^{-2}\textrm{ cm}^2\textrm{ g}^{-1}$ \citep{Freedman2008} to a dusty $\kappa=10^{-1}\textrm{ cm}^2\textrm{ g}^{-1}$. In terms of its treatment of the temperature behind the accretion shock, the model is compatible with hot start evolutionary models \citep{Fortney2005,Fortney2008,Marley2007}. 

By construction in the \citet{GinzburgChiang2019} model, the planet's accretion rate is given by $\dot{M}\sim M/t$, where $M$ is the planet's mass and $t$ is the system's age; this equality is naturally satisfied if accretion is regulated by a gap. With this assumption, a measured luminosity $L=GM\dot{M}/R$ and an estimated age $t$ can be mapped to $M$ and $\dot{M}$ using figure 7 in \citet{GinzburgChiang2019}. The planet's radius $R(M,\dot{M})$ is given by figures 5 and 6 of that paper. We emphasize that $\dot{M}=M/t$ in these figures is the average accretion rate. We discuss the translation to an instantaneous rate below.

\begin{figure*}
    \centering
    \includegraphics[width=\textwidth]{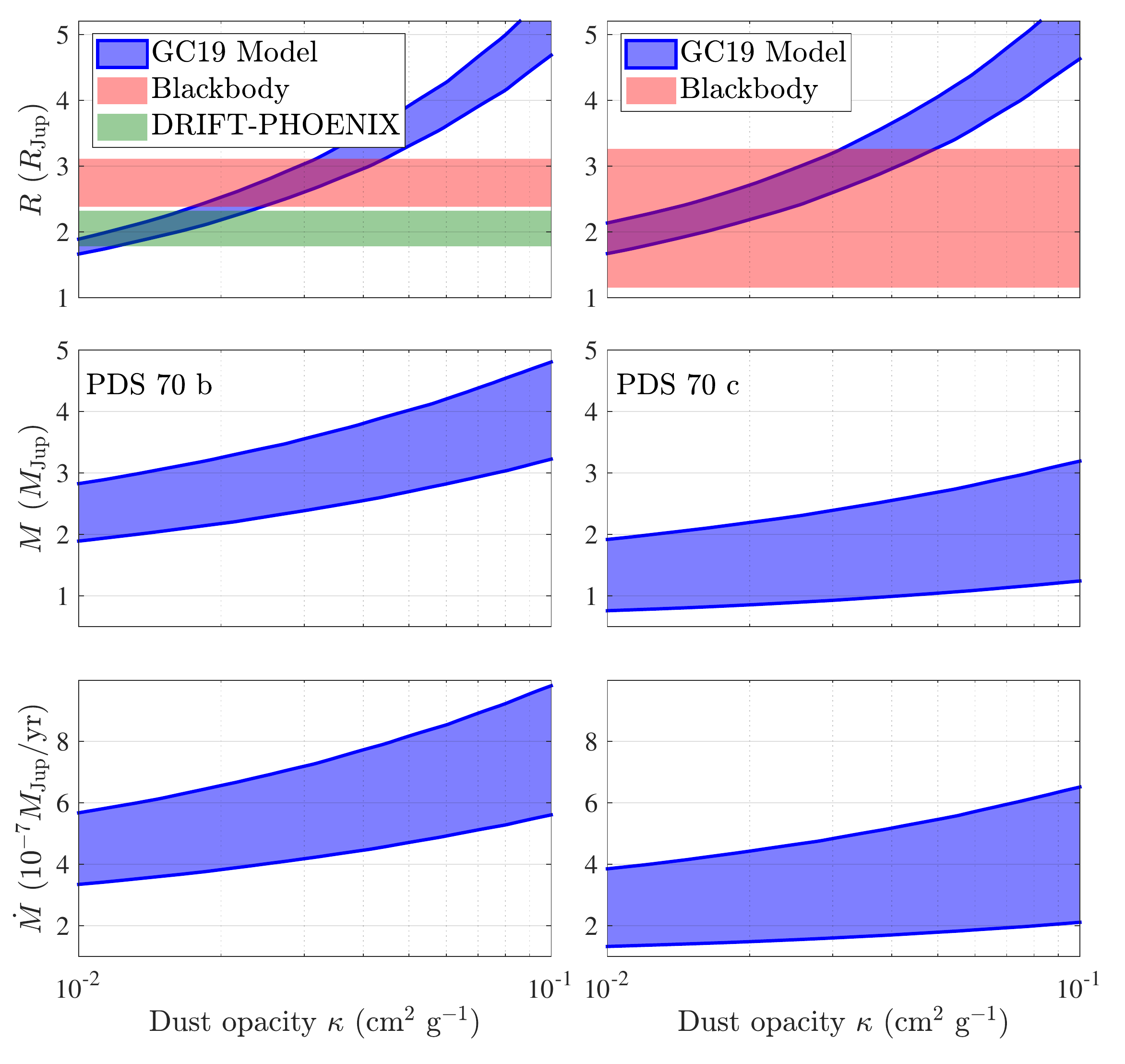}
    \caption{Radius, mass, and average accretion rate (blue diagonal stripes) of PDS 70 b (left panels) and PDS 70 c (right panels), as inferred by the evolutionary model of \citet{GinzburgChiang2019}. The model inputs are the system's age $t=5.4\pm 1.0$ Myr \citep{Muller2018} and bolometric luminosities $L_b=1.48^{+0.58}_{-0.30} \times 10^{-4} L_\odot$ and $L_c=3.60^{+5.84}_{-1.93} \times 10^{-5} L_\odot$, where we have averaged the luminosities inferred from the blackbody, two-blackbody, and DRIFT-PHOENIX models. The planet's radius at a given age is a function of the opacity $\kappa$ at its radiative--convective boundary, with high-opacity, dusty atmospheres contracting slower than low-opacity, dust-free ones. We use the radius estimates from the different atmospheric models (red and green horizontal stripes) to constrain $\kappa$. For PDS 70 b we do not include the two-blackbody radius range because it is spanned by the other two models (Table \ref{table:atm_models}). Similarly, for PDS 70 c we include only the blackbody model, as it spans the range of the other models (Table \ref{table:c_atm_models}; we exclude radii below $1~R_{\rm jup}$ due to electron degeneracy, assuming a roughly solar composition). Note that the instantaneous accretion rate is lower than the average rate ($M/t$) presented here by a factor of a few (see Section \ref{sec:evo_models}).}
    \label{fig:evol}
\end{figure*}

We plot our results in Figure \ref{fig:evol}. The radii, masses, and average accretion rates of PDS 70 b and c are inferred from their bolometric luminosities (average of the blackbody, two-blackbody, and DRIFT-PHOENIX models) and the estimated age of the system \citep{Muller2018}. We also compare the theoretically inferred radii from the evolutionary model to the SED constraints (horizontal red and green stripes for the different atmospheric models). The joint constraints on the radius of PDS 70 b imply that $0.01\lesssim \kappa\lesssim 0.04\textrm{ cm}^2\textrm{ g}^{-1}$, $2\lesssim M_b\lesssim 4~M_{\rm Jup}$, and $3 \times 10^{-7} \lesssim \dot{M}_b \lesssim 8 \times 10^{-7}~M_{\textrm{Jup}}\textrm{ yr}^{-1}$. The atmospheric models are less constraining for the radius of PDS 70 c. If we assume similar opacities for both planets, then $1\lesssim M_c \lesssim 3~M_{\rm Jup}$ and $1 \times 10^{-7} \lesssim \dot{M}_c \lesssim 5 \times~10^{-7} M_{\textrm{Jup}}\textrm{ yr}^{-1}$. This implies that the planets are two of the lowest mass directly-imaged planets. 
The mass accretion rates are consistent (by construction in the model) with the conclusion found in previous works that the planets are near the end of their formation process. 
We note that the largest uncertainty in inferring the planet's radius from its luminosity using this evolutionary model is due to the error in the age estimate. As seen in figures 5 and 6 of \citet{GinzburgChiang2019}, the radius at a few Myrs is mainly a function of age, almost independently of the planet's mass, accretion rate, and therefore luminosity.

Previous mass estimates for the PDS 70 planets have generally relied on either hot-start evolutionary models \citep{Baraffe2003} or deriving the mass from the surface gravity of the atmospheric fit. \citet{Muller2018} found a mass $2<M_b<17$ $M_{\rm Jup}$ using the radius and $\log g$ inferred from atmospheric models, whereas \citet{Keppler2018} found a narrower range $5<M_b<9$ $M_{\rm Jup}$ by comparing the \textit{H}, \textit{K}, and \textit{L} colors and magnitudes to hot start evolutionary models of fully formed planets at the age of the system. These values are a factor of 2 higher than our mass estimate for PDS 70 b. 
\citet{Haffert2019} used a similar comparison of \textit{K}-\textit{L} colors and \textit{L} magnitudes to hot start evolutionary models to estimate $4<M_c<12$ $M_{\rm Jup}$, also higher than our estimate for PDS 70 c. Again, this estimate relies on models that assume fully formed planets radiating away heat in isolation.
\citet{Christiaens2019APJ} estimated a mass $M_b\approx 2~M_{\rm Jup}$ in their isolated planet atmospheric models when fitting $\log g$ and the radius, similar to our lower limit. However, when they added a circumplanetary disk to their model they found a mass of $M_b\approx 10~M_{\rm Jup}$, which is above our upper limit.
More recently, \citet{Hashimoto2020} measured the width of the H$\alpha$ line to estimate masses of  $12 \pm 3~ M_{\rm Jup}$ and $11 \pm 5~ M_{\rm Jup}$ for PDS 70 b and c respectively. 
We note that their mass estimate depends on the square of the free-fall velocity, which is hard to measure directly, and was instead estimated using the accretion-shock model of \citet{Aoyama2018} and the assumption that the H$\alpha$ lines they measured were broadened beyond the instrumental resolution.

Our inferred masses could be too low if our SED fits significantly underestimate the total luminosity of the planets. Quantitatively, the inferred masses in the \citet{GinzburgChiang2019} evolutionary model are roughly proportional to the square root of the total luminosity.
As the current infrared data only reaches out to 4~$\mu$m, there could be emission at longer wavelengths that is unaccounted for, as a larger mid- to far-infrared peak in the SED is predicted in circumplanetary disk models \citep{Zhu2015,Szulagyi2019}. Indeed, \citet{Isella2019} detected emission from the planets at 855~$\mu$m with ALMA and interpreted the emission as coming from circumplanetary material. The PDS 70 b detection is not coincident with the planet (it is over 60~mas away from our orbit predictions), so we do not consider it as coming directly from the planet or its Hill sphere ($<$ 20~mas in radius). The PDS 70 c detection is consistent with our orbit prediction, and is a SNR$\approx$5 detection, indicating it is robust. The $106 \pm 19~\mu$Jy mm flux is a factor of $\sim$100 higher than what is predicted from our blackbody or two-blackbody fits. As a result, if we try to fit a two-blackbody model that includes this ALMA point, we find luminosities up to 100 times higher, requiring the planet to be $\gtrsim$10~$M_{\rm Jup}$. However, the dominant source of energy powering this emission does not have to be from the accreting planet. \citet{Isella2019} calculated that the equilibrium temperature of circumplanetary dust at the location of PDS 70 c to be 80~K, and that reprocessed stellar radiation is the dominant energy source if circumplanetary material fills up a significant fraction of its Hill sphere. Thus, the ALMA detection of PDS 70 c could be dominated by the re-radiation of starlight just as how the circumstellar disk is detected at these wavelengths. If this energy is not driven by planetary accretion, then it is not part of the energy balance of accretion that is at the foundation of the \citet{GinzburgChiang2019} model and thus should not be considered in estimating the mass and mass accretion rate. However, even if the emission can be fully explained by stellar heating, part of the mm flux could be due to planetary accretion, which would drive up the inferred masses presented in this work. Better constraints on the SED and in particular longer wavelength data are necessary to disentangle these effects. 

Previous accretion rate estimates have relied on hydrogen emission lines and primarily the H$\alpha$ line. \citet{Wagner2018} estimated the accretion rate onto PDS 70 b by converting the H$\alpha$ luminosity into an accretion luminosity. This conversion is poorly calibrated for planetary mass objects and potentially suffers from a large scatter \citep{Rigliaco2012,AoyamaIkoma2019,Thanathibodee2019}. This luminosity is then used to calculate $\dot{M}$ by adopting the mass range from hot start evolutionary models and assuming a planet radius equivalent to that of Jupiter. In our model, by contrast, the radius is calculated self-consistently using an evolutionary model appropriate for accreting planets. \citet{Wagner2018} state an upper limit of $\dot{M}_b<10^{-7}M_{\rm Jup}\textrm{ yr}^{-1}$, about 4 times lower than our estimate. \citet{Haffert2019} use the width of the H$\alpha$ line to infer mass accretion rates for both PDS 70 b and PDS 70 c, as it is independent of extinction. However, such a model is calibrated on higher mass brown dwarfs that form in isolation, and was noted to have large uncertainties for individual objects \citep{Natta2004}. The mass accretion rates reported in \citet{Haffert2019} are about a factor of 10 lower than what we find in this work.
\citet{AoyamaIkoma2019} model H$\alpha$ emission from the accretion shock and estimate $10^{-8}M_{\rm Jup}\textrm{ yr}^{-1}<\dot{M}_b<10^{-7}M_{\rm Jup}\textrm{ yr}^{-1}$ and $\dot{M}_c\sim 10^{-8}M_{\rm Jup}\textrm{ yr}^{-1}$. The higher accretion rates found in our work compared to all of these H$\alpha$ derived accretion rates can be partly explained by the difference between the mean and instantaneous accretion rates if the accretion rate gradually decreases over time, as we discuss below. Using the same model as \citet{AoyamaIkoma2019}, \citet{Hashimoto2020} estimated $\dot{M}_b>5\times 10^{-7}M_{\rm Jup}\textrm{ yr}^{-1}$ and $\dot{M}_c>1 \times 10^{-7}M_{\rm Jup}\textrm{ yr}^{-1}$ by combining their H$\alpha$ emission and upper limits on H$\beta$ emission to place lower limits on extinction. While this is consistent with our rate estimate, their inferred planet masses are significantly higher than ours as we discussed above.
\citet{Christiaens2019MNRAS} set an upper limit from the non-detection of Br$\gamma$ emission of $\dot{M}_b < 1.26 \times 10^{-7} (5 M_{\rm{Jup}}/M_b) (R_b/R_{\rm{Jup}}) M_{\rm Jup}\textrm{ yr}^{-1}$; this limit is consistent with our $\dot{M}_b$ values, given our $M$ and $R$. 

One significant difference is that our model calculates the mean accretion rate while the hydrogen emission lines are related to the instantaneous accretion rate. Our inferred mass accretion rates agree better with previous estimates if we refine the assumption that $\dot{M}=M/t$. More precisely, gap opening theory predicts that the planet's mass grows as $M\propto t^{1/\beta}$ up to the dispersal of the nebula, so $\dot{M}=\beta^{-1}M/t$. \citet{GinzburgChiang2019} consider two cases: $\beta=3$ for gaps opened in viscous disks and $\beta\approx 15$ for low-viscosity ones. \citet{TanigawaTanaka16}, on the other hand, suggest $\beta=5/3$ for gap-limited accretion (their equation 12). A different scenario, in which the planet's growth is limited by a roughly constant viscous transport rate across the disk (rather than by the gap), can be modeled with $\beta=1$. We conclude that coefficients $\beta>1$ may reconcile our estimate with the somewhat lower values found by other methods. Since $\beta$ is model dependent, we present the more robust average accretion rate $M/t$, and keep in mind that the instantaneous rate can be lower by a factor of few. 

\subsection{The Dusty Atmospheres of PDS 70 b and c}

The emission spectra of most directly imaged planets and brown dwarfs with temperatures similar to those inferred for PDS 70 b and c show extensive features from 1--5 $\mu$m caused by methane and water absorption  \citep[e.g.][]{Liu2013,filippazzo2015,bonnefoy2016,rajan2017}. These features are especially prominent due to the loss of mineral cloud opacity to cloud break up \citep{marley2010} and/or the sinking of the clouds below the photosphere \citep{stephens2009}. By comparison, the similarity of PDS 70 b and c's SEDs to blackbodies and their extreme redness in J-K in comparison to other objects \citep{Mesa2019} suggest much more dusty atmospheres. One possible explanation is the persistence of mineral clouds in the atmospheres of PDS 70 b and c despite their low temperatures due to their low gravities, as has been hypothesized to explain the redness of other low-gravity objects \citep[e.g.][]{Barman2011,Marley2012,liu2016}. 

Alternatively, the dust could stem from accretion itself, either via direct delivery of solids or from the recondensation of refractory material vaporized in the accretion shock. Assuming a 1:100 mass ratio of refractory to volatile material, the inferred planetary radii from the fits in Section \ref{sec:sed}, and the mass accretion rates from Section \ref{sec:evo_models}, the flux of refractory material to the planet is $\sim$10$^{-8}$--10$^{-7}$ g cm$^{-2}$ s$^{-1}$ for both PDS 70 b and c. We can compare this flux to the maximum column mass density of mineral clouds, $M_{cld}$, to evaluate the importance of accreted dust to the total atmospheric dust opacity. Assuming that the mineral clouds are composed entirely of forsterite and that the cloud material is well-mixed throughout the atmosphere above the cloud base, we approximate $M_{cld}$ as,

\begin{equation}
    M_{cld} \sim  f_{Mg} \frac{\mu_{For}}{2\mu_a} \rho_{cb} H =  f_{Mg} \frac{\mu_{For}}{2\mu_a} \frac{P_{cb}}{g}
\end{equation}

\noindent where $f_{Mg}$ is the mole fraction of magnesium, the limiting element in forsterite; $\mu_{For}$ and $\mu_a$ are the molecular weights of forsterite and the atmosphere, respectively; $\rho_{cb}$ and $P_{cb}$ are the density and pressure at the cloud base, respectively; $H$ is the scale height; and $g$ is the gravitational acceleration. Assuming solar abundances and a cloud base at 1 bar \citep{burrows1997}, we find column masses of mineral (forsterite) clouds $\sim$2 and $\sim$1.5 g cm$^{-2}$ for PDS 70 b and c, respectively. Dividing these values by the accreted dust flux yields timescales of $\sim$1 year, which is the time needed for the column mass density of the accreted dust to build up to that of the mineral clouds, assuming that the clouds and dust are both well-mixed and that the accreted dust has no sink. In actuality, most of the cloud mass will be concentrated near the cloud base \citep{gao2018}, allowing for accreted dust to potentially dominate the opacity at lower pressures. On the other hand, the accreted dust will be readily lost to evaporation once transported to higher pressures, e.g. below the cloud base. 

The transportation timescale of the dust is a strong function of dust particle size and the mixing timescale of the atmosphere. For example, for particles with radii of 1 $\mu$m, the sedimentation timescale over 1 scale height near the cloud base is $\sim$20 years. If we parameterize the mixing with eddy diffusion, then the mixing timescale over the thickness of the atmosphere from the homopause to 1 bar is $\sim$10 years for an eddy diffusion coefficient of 10$^8$ cm$^2$ s$^{-1}$ \citep{moses2016}. Both of these timescales suggest that accreted dust could be a dominant opacity source in the atmospheres of PDS 70 b and c, since they are longer than the $\sim$1 year it takes for dust opacity to build up to match that of the clouds. This accreted dust would mask deep molecular absorption features seen in more mature giant planets, though the accretion of volatile materials could still produce some signatures. Dust contribution to the opacity is also consistent with the relatively large radius inferred for PDS 70 b (see Fig. \ref{fig:evol}).
While we did not explore more detailed atmosphere models, the youth of the planets, the observed H$\alpha$ emission, and the gas-rich circumstellar environment are all consistent with a dusty accreting planet hypothesis. Some emission could also be coming from circumplanetary material, but the available observations are sufficiently fit by a single component blackbody. 

\section{Conclusion}

We present new thermal \Lp{}-band imaging of the PDS 70 planetary system with Keck/NIRC2 and the new infrared pyramid wavefront sensor as part of its science verification. After modeling out the circumstellar disk, we detected both PDS 70 b and c and measured their astrometry and \Lp{}-band photometry. The orbits are still relatively unconstrained, so we constructed physically motivated priors to estimate $a_b = 20^{+3}_{-4}$~au and $a_c = 34^{+12}_{-6}$~au. These orbits are within $\sim20^\circ$ of being coplanar with the circumstellar disk. 

We find that our \Lp{}-band photometry helps constrain the total luminosity and radius of PDS 70 b by placing more precise bounds on the red half of its SED. We find a radius for the photosphere between 2-3~$R_{Jup}$. The SED of PDS 70 c is still relatively unconstrained, but we can constrain the total luminosity of PDS 70 c to within an order of magnitude. While it is still unclear what models can accurately describe the SED of either planet, we found that a single blackbody SED had the most empirical support out of the four models we considered. More data is needed to warrant fitting more sophisticated models to the data.

With the inferred luminosities and radii of PDS 70 b and c, we used the evolutionary model of \citet{GinzburgChiang2019} for accreting protoplanets to constrain the mass and mass accretion rate of these two planets. We find a mass of PDS 70 b between 2 and 4 $M_{\textrm{Jup}}$ and a mean mass accretion rate between $3 \times 10^{-7}$ and $8 \times 10^{-7}~M_{\textrm{Jup}}\textrm{ yr}^{-1}$. For PDS 70 c, we find a mass between 1 and 3 $M_{\textrm{Jup}}$ and mean mass accretion rate between $1 \times 10^{-7}$ and $5 \times 10^{-7} M_{\textrm{Jup}}\textrm{ yr}^{-1}$. The instantaneous rates are lower by a factor of a few, which depends on the specifics of the accretion model.
The mass estimates make PDS 70 b and c two of the lowest mass directly-imaged planets to date. The mass accretion rates imply dust accretion timescales short enough to shroud both planets, consistent with the absence of strong molecular absorption features in their SEDs.

\acknowledgments
We thank Trevor David for helpful discussion and the referee for their helpful suggestions to the paper. J.J.W., S.G., and P.G. are supported by the Heising-Simons Foundation 51 Pegasi b postdoctoral fellowship. 
The Keck infrared pyramid wavefront sensor was developed with support from the National Science Foundation under grants AST-1611623 and AST-1106391, as well as the Heising Simons Foundation under the Keck Planet Imager and Characterizer project.
This research is partially supported by NASA ROSES XRP, award 80NSSC19K0294.
FMe acknowledges funding from ANR of France under contract ANR-16-CE31-0013.
This project has received funding from the European Research Council (ERC) under the European Union's Seventh Framework Program (grant agreement 337569, VORTEX) and under the European Union's Horizon 2020 research and innovation programme (grant agreement 819155, EPIC). The research was supported by the Wallonia-Brussels Federation (grant for Concerted Research Actions).
Part of the computations presented here were conducted on the Caltech High Performance Cluster, partially supported by a grant from the Gordon and Betty Moore Foundation.
Data presented in this work were obtained at the W. M. Keck Observatory, which is operated as a scientific partnership among the California Institute of Technology, the University of California and the National Aeronautics and Space Administration. The Observatory was made possible by the generous financial support of the W. M. Keck Foundation. We wish to recognize and acknowledge the very significant cultural role and reverence that the summit of Maunakea has always had within the indigenous Hawaiian community. We are most fortunate to have the opportunity to conduct observations from this mountain.

\facility{Keck II (NIRC2)}

\software{{\tt pyKLIP} \citep{Wang2015}, {\tt orbitize!} \citep{Blunt2020}, {\tt DebrisDiskFM} \citep{Ren2019}, {\tt emcee} \citep{ForemanMackey2013}, {\tt ptemcee} \citep{Vousden2016}, {\tt MCFOST} \citep{Pinte2006,Pinte2009} }


\bibliography{refs}

\begin{thebibliography}{}
\expandafter\ifx\csname natexlab\endcsname\relax\def\natexlab#1{#1}\fi
\providecommand{\url}[1]{\href{#1}{#1}}
\providecommand{\dodoi}[1]{}
\providecommand{\doarXiv}[1]{\href{https://arxiv.org/abs/#1}{\nolinkurl{https://arxiv.org/abs/#1}}}

\bibitem[{Akaike {et~al.}(1973)Akaike, Petrov, \& Csaki}]{Akaike1973}
Akaike, H., Petrov, B.~N., \& Csaki, F. 1973, Second international symposium on
  information theory,  Akad{\'e}miai Kiad{\'o}, Budapest

\bibitem[{{Allard} {et~al.}(2012){Allard}, {Homeier}, \&
  {Freytag}}]{Allard2012}
{Allard}, F., {Homeier}, D., \& {Freytag}, B. 2012,
  \href{http://dx.doi.org/10.1098/rsta.2011.0269}{\color{magenta}RSPTA},
  \href{https://ui.adsabs.harvard.edu/abs/2012RSPTA.370.2765A}{\color{blue}370},
  \href{https://ui.adsabs.harvard.edu/abs/2012RSPTA.370.2765A}{\color{blue}2765}

\bibitem[{{Aoyama} \& {Ikoma}(2019)}]{AoyamaIkoma2019}
{Aoyama}, Y., \& {Ikoma}, M. 2019,
  \href{http://dx.doi.org/10.3847/2041-8213/ab5062}{\color{magenta}\apjl},
  \href{https://ui.adsabs.harvard.edu/abs/2019ApJ...885L..29A}{\color{blue}885},
  \href{https://ui.adsabs.harvard.edu/abs/2019ApJ...885L..29A}{\color{blue}L29}

\bibitem[{{Aoyama} {et~al.}(2018){Aoyama}, {Ikoma}, \& {Tanigawa}}]{Aoyama2018}
{Aoyama}, Y., {Ikoma}, M., \& {Tanigawa}, T. 2018,
  \href{http://dx.doi.org/10.3847/1538-4357/aadc11}{\color{magenta}\apj},
  \href{https://ui.adsabs.harvard.edu/abs/2018ApJ...866...84A}{\color{blue}866},
  \href{https://ui.adsabs.harvard.edu/abs/2018ApJ...866...84A}{\color{blue}84}

\bibitem[{{Augereau} {et~al.}(1999){Augereau}, {Lagrange}, {Mouillet},
  {Papaloizou}, \& {Grorod}}]{Augereau1999}
{Augereau}, J.~C., {Lagrange}, A.~M., {Mouillet}, D., {et~al.} 1999, \aap,
  \href{http://ui.adsabs.harvard.edu/abs/1999A%26A...348..557A}{\color{blue}348},
  \href{http://ui.adsabs.harvard.edu/abs/1999A%26A...348..557A}{\color{blue}557}

\bibitem[{{Baraffe} {et~al.}(2003){Baraffe}, {Chabrier}, {Barman}, {Allard}, \&
  {Hauschildt}}]{Baraffe2003}
{Baraffe}, I., {Chabrier}, G., {Barman}, T.~S., {et~al.} 2003,
  \href{http://dx.doi.org/10.1051/0004-6361:20030252}{\color{magenta}\aap},
  \href{https://ui.adsabs.harvard.edu/abs/2003A&A...402..701B}{\color{blue}402},
  \href{https://ui.adsabs.harvard.edu/abs/2003A&A...402..701B}{\color{blue}701}

\bibitem[{{Barman} {et~al.}(2011){Barman}, {Macintosh}, {Konopacky}, \&
  {Marois}}]{Barman2011}
{Barman}, T.~S., {Macintosh}, B., {Konopacky}, Q.~M., \& {Marois}, C. 2011,
  \href{http://dx.doi.org/10.1088/0004-637X/733/1/65}{\color{magenta}\apj},
  \href{https://ui.adsabs.harvard.edu/abs/2011ApJ...733...65B}{\color{blue}733},
  \href{https://ui.adsabs.harvard.edu/abs/2011ApJ...733...65B}{\color{blue}65}

\bibitem[{{Bean} \& {Seifahrt}(2009)}]{Bean2009}
{Bean}, J.~L., \& {Seifahrt}, A. 2009,
  \href{http://dx.doi.org/10.1051/0004-6361/200811280}{\color{magenta}\aap},
  \href{https://ui.adsabs.harvard.edu/abs/2009A&A...496..249B}{\color{blue}496},
  \href{https://ui.adsabs.harvard.edu/abs/2009A&A...496..249B}{\color{blue}249}

\bibitem[{{Blunt} {et~al.}(2020){Blunt}, {Wang}, {Angelo}, {Ngo}, {Cody}, {De
  Rosa}, {Graham}, {Hirsch}, {Nagpal}, {Nielsen}, {Pearce}, {Rice}, \&
  {Tejada}}]{Blunt2020}
{Blunt}, S., {Wang}, J.~J., {Angelo}, I., {et~al.} 2020,
  \href{http://dx.doi.org/10.3847/1538-3881/ab6663}{\color{magenta}\aj},
  \href{https://ui.adsabs.harvard.edu/abs/2020AJ....159...89B}{\color{blue}159},
  \href{https://ui.adsabs.harvard.edu/abs/2020AJ....159...89B}{\color{blue}89}

\bibitem[{{Bodenheimer}(1974)}]{Bodenheimer1974}
{Bodenheimer}, P. 1974,
  \href{http://dx.doi.org/10.1016/0019-1035(74)90050-5}{\color{magenta}\icarus},
  \href{https://ui.adsabs.harvard.edu/abs/1974Icar...23..319B}{\color{blue}23},
  \href{https://ui.adsabs.harvard.edu/abs/1974Icar...23..319B}{\color{blue}319}

\bibitem[{{Bond} {et~al.}(2019){Bond}, {Cetre}, {Ragland}, {Lilley},
  {Wizinowich}, {Mawet}, {Chun}, {Delorme}, {Jovanovic}, {Wetherell},
  {Jacobson}, {Lockhart}, {Warmbier}, {Wallace}, {Hall}, {Goebel}, \&
  {Guyon}}]{Bond2019}
{Bond}, C., {Cetre}, S., {Ragland}, S., {et~al.} 2019, Adaptive Optics for
  Extremely Large Telescopes VI (AO4ELT6)

\bibitem[{{Bond} {et~al.}(2018){Bond}, {Wizinowich}, {Chun}, {Mawet}, {Lilley},
  {Cetre}, {Jovanovic}, {Delorme}, {Wetherell}, {Jacobson}, {Lockhart},
  {Warmbier}, {Wallace}, {Hall}, {Goebel}, {Guyon}, {Plantet}, {Agapito},
  {Giordano}, {Esposito}, \& {Femenia-Castella}}]{Bond2018}
{Bond}, C.~Z., {Wizinowich}, P., {Chun}, M., {et~al.} 2018,
  \href{http://dx.doi.org/10.1117/12.2314121}{\color{magenta}Proc.~SPIE},
  \href{https://ui.adsabs.harvard.edu/abs/2018SPIE10703E..1ZB}{\color{blue}10703},
  \href{https://ui.adsabs.harvard.edu/abs/2018SPIE10703E..1ZB}{\color{blue}107031Z}

\bibitem[{{Bonnefoy} {et~al.}(2016){Bonnefoy}, {Zurlo}, {Baudino}, {Lucas},
  {Mesa}, {Maire}, {Vigan}, {Galicher}, {Homeier}, {Marocco}, {Gratton},
  {Chauvin}, {Allard}, {Desidera}, {Kasper}, {Moutou}, {Lagrange}, {Antichi},
  {Baruffolo}, {Baudrand }, {Beuzit}, {Boccaletti}, {Cantalloube}, {Carbillet},
  {Charton}, {Claudi}, {Costille}, {Dohlen}, {Dominik}, {Fantinel},
  {Feautrier}, {Feldt}, {Fusco}, {Gigan}, {Girard}, {Gluck}, {Gry}, {Henning},
  {Janson}, {Langlois}, {Madec}, {Magnard}, {Maurel}, {Mawet}, {Meyer},
  {Milli}, {Moeller-Nilsson}, {Mouillet}, {Pavlov}, {Perret}, {Pujet}, {Quanz},
  {Rochat}, {Rousset}, {Roux}, {Salasnich}, {Salter}, {Sauvage}, {Schmid},
  {Sevin}, {Soenke}, {Stadler}, {Turatto}, {Udry}, {Vakili}, {Wahhaj}, \&
  {Wildi}}]{bonnefoy2016}
{Bonnefoy}, M., {Zurlo}, A., {Baudino}, J.~L., {et~al.} 2016,
  \href{http://dx.doi.org/10.1051/0004-6361/201526906}{\color{magenta}\aap},
  \href{https://ui.adsabs.harvard.edu/abs/2016A&A...587A..58B}{\color{blue}587},
  \href{https://ui.adsabs.harvard.edu/abs/2016A&A...587A..58B}{\color{blue}A58}

\bibitem[{{Boss}(1998)}]{Boss1998}
{Boss}, A.~P. 1998,
  \href{http://dx.doi.org/10.1086/306036}{\color{magenta}\apj},
  \href{https://ui.adsabs.harvard.edu/abs/1998ApJ...503..923B}{\color{blue}503},
  \href{https://ui.adsabs.harvard.edu/abs/1998ApJ...503..923B}{\color{blue}923}

\bibitem[{Burnham \& Anderson(2002)}]{Burnham}
Burnham, K., \& Anderson, D. 2002, Model Selection and Multimodel Inference: A
  Practical Information-Theoretic Approach, 2nd edn. (Springer-Verlag New York)

\bibitem[{{Burrows} {et~al.}(1997){Burrows}, {Marley}, {Hubbard}, {Lunine},
  {Guillot}, {Saumon}, {Freedman}, {Sudarsky}, \& {Sharp}}]{burrows1997}
{Burrows}, A., {Marley}, M., {Hubbard}, W.~B., {et~al.} 1997,
  \href{http://dx.doi.org/10.1086/305002}{\color{magenta}\apj},
  \href{https://ui.adsabs.harvard.edu/abs/1997ApJ...491..856B}{\color{blue}491},
  \href{https://ui.adsabs.harvard.edu/abs/1997ApJ...491..856B}{\color{blue}856}

\bibitem[{{Choi} {et~al.}(2016){Choi}, {Dotter}, {Conroy}, {Cantiello},
  {Paxton}, \& {Johnson}}]{Choi:2016kf}
{Choi}, J., {Dotter}, A., {Conroy}, C., {et~al.} 2016,
  \href{http://dx.doi.org/10.3847/0004-637X/823/2/102}{\color{magenta}\apj},
  \href{https://ui.adsabs.harvard.edu/abs/2016ApJ...823..102C}{\color{blue}823},
  \href{https://ui.adsabs.harvard.edu/abs/2016ApJ...823..102C}{\color{blue}102}

\bibitem[{{Christiaens} {et~al.}(2019{\natexlab{a}}){Christiaens},
  {Cantalloube}, {Casassus}, {Price}, {Absil}, {Pinte}, {Girard}, \&
  {Montesinos}}]{Christiaens2019APJ}
{Christiaens}, V., {Cantalloube}, F., {Casassus}, S., {et~al.}
  2019{\natexlab{a}},
  \href{http://dx.doi.org/10.3847/2041-8213/ab212b}{\color{magenta}\apjl},
  \href{https://ui.adsabs.harvard.edu/abs/2019ApJ...877L..33C}{\color{blue}877},
  \href{https://ui.adsabs.harvard.edu/abs/2019ApJ...877L..33C}{\color{blue}L33}

\bibitem[{{Christiaens} {et~al.}(2019{\natexlab{b}}){Christiaens}, {Casassus},
  {Absil}, {Cantalloube}, {Gomez Gonzalez}, {Girard}, {Ram{\'\i}rez}, {Pairet},
  {Salinas}, {Price}, {Pinte}, {Quanz}, {Jord{\'a}n}, {Mawet}, \&
  {Wahhaj}}]{Christiaens2019MNRAS}
{Christiaens}, V., {Casassus}, S., {Absil}, O., {et~al.} 2019{\natexlab{b}},
  \href{http://dx.doi.org/10.1093/mnras/stz1232}{\color{magenta}\mnras},
  \href{https://ui.adsabs.harvard.edu/abs/2019MNRAS.486.5819C}{\color{blue}486},
  \href{https://ui.adsabs.harvard.edu/abs/2019MNRAS.486.5819C}{\color{blue}5819}

\bibitem[{{Currie} {et~al.}(2015){Currie}, {Cloutier}, {Brittain}, {Grady},
  {Burrows}, {Muto}, {Kenyon}, \& {Kuchner}}]{Currie2015}
{Currie}, T., {Cloutier}, R., {Brittain}, S., {et~al.} 2015,
  \href{http://dx.doi.org/10.1088/2041-8205/814/2/L27}{\color{magenta}\apjl},
  \href{https://ui.adsabs.harvard.edu/abs/2015ApJ...814L..27C}{\color{blue}814},
  \href{https://ui.adsabs.harvard.edu/abs/2015ApJ...814L..27C}{\color{blue}L27}

\bibitem[{{Currie} {et~al.}(2018){Currie}, {Brandt}, {Uyama}, {Nielsen},
  {Blunt}, {Guyon}, {Tamura}, {Marois}, {Mede}, {Kuzuhara}, {Groff},
  {Jovanovic}, {Kasdin}, {Lozi}, {Hodapp}, {Chilcote}, {Carson}, {Martinache},
  {Goebel}, {Grady}, {McElwain}, {Akiyama}, {Asensio-Torres}, {Hayashi},
  {Janson}, {Knapp}, {Kwon}, {Nishikawa}, {Oh}, {Schlieder}, {Serabyn},
  {Sitko}, \& {Skaf}}]{Currie2018}
{Currie}, T., {Brandt}, T.~D., {Uyama}, T., {et~al.} 2018,
  \href{http://dx.doi.org/10.3847/1538-3881/aae9ea}{\color{magenta}\aj},
  \href{https://ui.adsabs.harvard.edu/abs/2018AJ....156..291C}{\color{blue}156},
  \href{https://ui.adsabs.harvard.edu/abs/2018AJ....156..291C}{\color{blue}291}

\bibitem[{{Cutri} {et~al.}(2013)}]{Cutri2013}
{Cutri}, R.~M., {et~al.} 2013, VizieR Online Data Catalog,
  \href{https://ui.adsabs.harvard.edu/abs/2013yCat.2328....0C}{\color{blue}II/328}

\bibitem[{{Czekala} {et~al.}(2015){Czekala}, {Andrews}, {Mandel}, {Hogg}, \&
  {Green}}]{Czekala2015}
{Czekala}, I., {Andrews}, S.~M., {Mandel}, K.~S., {et~al.} 2015,
  \href{http://dx.doi.org/10.1088/0004-637X/812/2/128}{\color{magenta}\apj},
  \href{https://ui.adsabs.harvard.edu/abs/2015ApJ...812..128C}{\color{blue}812},
  \href{https://ui.adsabs.harvard.edu/abs/2015ApJ...812..128C}{\color{blue}128}

\bibitem[{{De Rosa} {et~al.}(2016){De Rosa}, {Rameau}, {Patience}, {Graham},
  {Doyon}, {Lafreni{\`e}re}, {Macintosh}, {Pueyo}, {Rajan}, {Wang},
  {Ward-Duong}, {Hung}, {Maire}, {Nielsen}, {Ammons}, {Bulger}, {Cardwell},
  {Chilcote}, {Galvez}, {Gerard}, {Goodsell}, {Hartung}, {Hibon}, {Ingraham},
  {Johnson-Groh}, {Kalas}, {Konopacky}, {Marchis}, {Marois}, {Metchev},
  {Morzinski}, {Oppenheimer}, {Perrin}, {Rantakyr{\"o}}, {Savransky}, \&
  {Thomas}}]{DeRosa2016}
{De Rosa}, R.~J., {Rameau}, J., {Patience}, J., {et~al.} 2016,
  \href{http://dx.doi.org/10.3847/0004-637X/824/2/121}{\color{magenta}\apj},
  \href{https://ui.adsabs.harvard.edu/abs/2016ApJ...824..121D}{\color{blue}824},
  \href{https://ui.adsabs.harvard.edu/abs/2016ApJ...824..121D}{\color{blue}121}

\bibitem[{{Dempsey} {et~al.}(2013){Dempsey}, {Friberg}, {Jenness}, {Tilanus},
  {Thomas}, {Holland}, {Bintley}, {Berry}, {Chapin}, {Chrysostomou}, {Davis},
  {Gibb}, {Parsons}, \& {Robson}}]{Dempsey2013}
{Dempsey}, J.~T., {Friberg}, P., {Jenness}, T., {et~al.} 2013,
  \href{http://dx.doi.org/10.1093/mnras/stt090}{\color{magenta}\mnras},
  \href{https://ui.adsabs.harvard.edu/abs/2013MNRAS.430.2534D}{\color{blue}430},
  \href{https://ui.adsabs.harvard.edu/abs/2013MNRAS.430.2534D}{\color{blue}2534}

\bibitem[{{Draine} \& {Lee}(1984)}]{draine84}
{Draine}, B.~T., \& {Lee}, H.~M. 1984,
  \href{http://dx.doi.org/10.1086/162480}{\color{magenta}\apj},
  \href{http://ads.bao.ac.cn/abs/1984ApJ...285...89D}{\color{blue}285},
  \href{http://ads.bao.ac.cn/abs/1984ApJ...285...89D}{\color{blue}89}

\bibitem[{{Esposito} {et~al.}(2018){Esposito}, {Duch{\^e}ne}, {Kalas}, {Rice},
  {Choquet}, {Ren}, {Perrin}, {Chen}, {Arriaga}, {Chiang}, {Nielsen}, {Graham},
  {Wang}, {De Rosa}, {Follette}, {Ammons}, {Ansdell}, {Bailey}, {Barman},
  {Sebasti{\'a}n Bruzzone}, {Bulger}, {Chilcote}, {Cotten}, {Doyon},
  {Fitzgerald}, {Goodsell}, {Greenbaum}, {Hibon}, {Hung}, {Ingraham},
  {Konopacky}, {Larkin}, {Macintosh}, {Maire}, {Marchis}, {Marois}, {Mazoyer},
  {Metchev}, {Millar-Blanchaer}, {Oppenheimer}, {Palmer}, {Patience},
  {Poyneer}, {Pueyo}, {Rajan}, {Rameau}, {Rantakyr{\"o}}, {Ryan}, {Savransky},
  {Schneider}, {Sivaramakrishnan}, {Song}, {Soummer}, {Thomas}, {Wallace},
  {Ward-Duong}, {Wiktorowicz}, \& {Wolff}}]{esposito18}
{Esposito}, T.~M., {Duch{\^e}ne}, G., {Kalas}, P., {et~al.} 2018,
  \href{http://dx.doi.org/10.3847/1538-3881/aacbc9}{\color{magenta}\aj},
  \href{http://ui.adsabs.harvard.edu/abs/2018AJ....156...47E}{\color{blue}156},
  \href{http://ui.adsabs.harvard.edu/abs/2018AJ....156...47E}{\color{blue}47}

\bibitem[{{Filippazzo} {et~al.}(2015){Filippazzo}, {Rice}, {Faherty}, {Cruz},
  {Van Gordon}, \& {Looper}}]{filippazzo2015}
{Filippazzo}, J.~C., {Rice}, E.~L., {Faherty}, J., {et~al.} 2015,
  \href{http://dx.doi.org/10.1088/0004-637X/810/2/158}{\color{magenta}\apj},
  \href{https://ui.adsabs.harvard.edu/abs/2015ApJ...810..158F}{\color{blue}810},
  \href{https://ui.adsabs.harvard.edu/abs/2015ApJ...810..158F}{\color{blue}158}

\bibitem[{{Follette} {et~al.}(2017){Follette}, {Rameau}, {Dong}, {Pueyo},
  {Close}, {Duch{\^e}ne}, {Fung}, {Leonard}, {Macintosh}, {Males}, {Marois},
  {Millar-Blanchaer}, {Morzinski}, {Mullen}, {Perrin}, {Spiro}, {Wang},
  {Ammons}, {Bailey}, {Barman}, {Bulger}, {Chilcote}, {Cotten}, {De Rosa},
  {Doyon}, {Fitzgerald}, {Goodsell}, {Graham}, {Greenbaum}, {Hibon}, {Hung},
  {Ingraham}, {Kalas}, {Konopacky}, {Larkin}, {Maire}, {Marchis}, {Metchev},
  {Nielsen}, {Oppenheimer}, {Palmer}, {Patience}, {Poyneer}, {Rajan},
  {Rantakyr{\"o}}, {Savransky}, {Schneider}, {Sivaramakrishnan}, {Song},
  {Soummer}, {Thomas}, {Vega}, {Wallace}, {Ward-Duong}, {Wiktorowicz}, \&
  {Wolff}}]{Follette2017}
{Follette}, K.~B., {Rameau}, J., {Dong}, R., {et~al.} 2017,
  \href{http://dx.doi.org/10.3847/1538-3881/aa6d85}{\color{magenta}\aj},
  \href{https://ui.adsabs.harvard.edu/abs/2017AJ....153..264F}{\color{blue}153},
  \href{https://ui.adsabs.harvard.edu/abs/2017AJ....153..264F}{\color{blue}264}

\bibitem[{{Foreman-Mackey} {et~al.}(2013){Foreman-Mackey}, {Hogg}, {Lang}, \&
  {Goodman}}]{ForemanMackey2013}
{Foreman-Mackey}, D., {Hogg}, D.~W., {Lang}, D., \& {Goodman}, J. 2013,
  \href{http://dx.doi.org/10.1086/670067}{\color{magenta}\pasp},
  \href{https://ui.adsabs.harvard.edu/abs/2013PASP..125..306F}{\color{blue}125},
  \href{https://ui.adsabs.harvard.edu/abs/2013PASP..125..306F}{\color{blue}306}

\bibitem[{{Fortney} {et~al.}(2005){Fortney}, {Marley}, {Hubickyj},
  {Bodenheimer}, \& {Lissauer}}]{Fortney2005}
{Fortney}, J.~J., {Marley}, M.~S., {Hubickyj}, O., {et~al.} 2005,
  \href{http://dx.doi.org/10.1002/asna.200510465}{\color{magenta}AN},
  \href{http://adsabs.harvard.edu/abs/2005AN....326..925F}{\color{blue}326},
  \href{http://adsabs.harvard.edu/abs/2005AN....326..925F}{\color{blue}925}

\bibitem[{{Fortney} {et~al.}(2008){Fortney}, {Marley}, {Saumon}, \&
  {Lodders}}]{Fortney2008}
{Fortney}, J.~J., {Marley}, M.~S., {Saumon}, D., \& {Lodders}, K. 2008,
  \href{http://dx.doi.org/10.1086/589942}{\color{magenta}\apj},
  \href{http://adsabs.harvard.edu/abs/2008ApJ...683.1104F}{\color{blue}683},
  \href{http://adsabs.harvard.edu/abs/2008ApJ...683.1104F}{\color{blue}1104}

\bibitem[{{Freedman} {et~al.}(2008){Freedman}, {Marley}, \&
  {Lodders}}]{Freedman2008}
{Freedman}, R.~S., {Marley}, M.~S., \& {Lodders}, K. 2008,
  \href{http://dx.doi.org/10.1086/521793}{\color{magenta}\apjs},
  \href{https://ui.adsabs.harvard.edu/abs/2008ApJS..174..504F}{\color{blue}174},
  \href{https://ui.adsabs.harvard.edu/abs/2008ApJS..174..504F}{\color{blue}504}

\bibitem[{{\textit{Gaia} Collaboration} {et~al.}(2018){\textit{Gaia}
  Collaboration}, {Brown}, {Vallenari}, {Prusti}, {de Bruijne}, {Babusiaux},
  {Bailer-Jones}, {Biermann}, {Evans}, {Eyer}, {Jansen}, {Jordi}, {Klioner},
  {Lammers}, {Lindegren}, {Luri}, {Mignard}, {Panem}, {Pourbaix}, {Randich},
  {Sartoretti}, {Siddiqui}, {Soubiran}, {van Leeuwen}, {Walton}, {Arenou},
  {Bastian}, {Cropper}, {Drimmel}, {Katz}, {Lattanzi}, {Bakker}, {Cacciari},
  {Casta{\~n}eda}, {Chaoul}, {Cheek}, {De Angeli}, {Fabricius}, {Guerra},
  {Holl}, {Masana}, {Messineo}, {Mowlavi}, {Nienartowicz}, {Panuzzo},
  {Portell}, {Riello}, {Seabroke}, {Tanga}, {Th{\'e}venin}, {Gracia-Abril},
  {Comoretto}, {Garcia-Reinaldos}, {Teyssier}, {Altmann}, {Andrae}, {Audard},
  {Bellas-Velidis}, {Benson}, {Berthier}, {Blomme}, {Burgess}, {Busso},
  {Carry}, {Cellino}, {Clementini}, {Clotet}, {Creevey}, {Davidson}, {De
  Ridder}, {Delchambre}, {Dell'Oro}, {Ducourant},
  {Fern{\'a}ndez-Hern{\'a}ndez}, {Fouesneau}, {Fr{\'e}mat}, {Galluccio},
  {Garc{\'\i}a-Torres}, {Gonz{\'a}lez-N{\'u}{\~n}ez}, {Gonz{\'a}lez-Vidal},
  {Gosset}, {Guy}, {Halbwachs}, {Hambly}, {Harrison}, {Hern{\'a}ndez},
  {Hestroffer}, {Hodgkin}, {Hutton}, {Jasniewicz}, {Jean-Antoine-Piccolo},
  {Jordan}, {Korn}, {Krone-Martins}, {Lanzafame}, {Lebzelter}, {L{\"o}ffler},
  {Manteiga}, {Marrese}, {Mart{\'\i}n-Fleitas}, {Moitinho}, {Mora}, {Muinonen},
  {Osinde}, {Pancino}, {Pauwels}, {Petit}, {Recio-Blanco}, {Richards},
  {Rimoldini}, {Robin}, {Sarro}, {Siopis}, {Smith}, {Sozzetti}, {S{\"u}veges},
  {Torra}, {van Reeven}, {Abbas}, {Abreu Aramburu}, {Accart}, {Aerts},
  {Altavilla}, {{\'A}lvarez}, {Alvarez}, {Alves}, {Anderson}, {Andrei},
  {Anglada Varela}, {Antiche}, {Antoja}, {Arcay}, {Astraatmadja}, {Bach},
  {Baker}, {Balaguer-N{\'u}{\~n}ez}, {Balm}, {Barache}, {Barata}, {Barbato},
  {Barblan}, {Barklem}, {Barrado}, {Barros}, {Barstow}, {Bartholom{\'e}
  Mu{\~n}oz}, {Bassilana}, {Becciani}, {Bellazzini}, {Berihuete}, {Bertone},
  {Bianchi}, {Bienaym{\'e}}, {Blanco-Cuaresma}, {Boch}, {Boeche}, {Bombrun},
  {Borrachero}, {Bossini}, {Bouquillon}, {Bourda}, {Bragaglia}, {Bramante},
  {Breddels}, {Bressan}, {Brouillet}, {Br{\"u}semeister}, {Brugaletta},
  {Bucciarelli}, {Burlacu}, {Busonero}, {Butkevich}, {Buzzi}, {Caffau},
  {Cancelliere}, {Cannizzaro}, {Cantat-Gaudin}, {Carballo}, {Carlucci},
  {Carrasco}, {Casamiquela}, {Castellani}, {Castro-Ginard}, {Charlot},
  {Chemin}, {Chiavassa}, {Cocozza}, {Costigan}, {Cowell}, {Crifo}, {Crosta},
  {Crowley}, {Cuypers}, {Dafonte}, {Damerdji}, {Dapergolas}, {David}, {David},
  {de Laverny}, {De Luise}, {De March}, {de Martino}, {de Souza}, {de Torres},
  {Debosscher}, {del Pozo}, {Delbo}, {Delgado}, {Delgado}, {Di Matteo},
  {Diakite}, {Diener}, {Distefano}, {Dolding}, {Drazinos}, {Dur{\'a}n},
  {Edvardsson}, {Enke}, {Eriksson}, {Esquej}, {Eynard Bontemps}, {Fabre},
  {Fabrizio}, {Faigler}, {Falc{\~a}o}, {Farr{\`a}s Casas}, {Federici},
  {Fedorets}, {Fernique}, {Figueras}, {Filippi}, {Findeisen}, {Fonti},
  {Fraile}, {Fraser}, {Fr{\'e}zouls}, {Gai}, {Galleti}, {Garabato},
  {Garc{\'\i}a-Sedano}, {Garofalo}, {Garralda}, {Gavel}, {Gavras}, {Gerssen},
  {Geyer}, {Giacobbe}, {Gilmore}, {Girona}, {Giuffrida}, {Glass}, {Gomes},
  {Granvik}, {Gueguen}, {Guerrier}, {Guiraud}, {Guti{\'e}rrez-S{\'a}nchez},
  {Haigron}, {Hatzidimitriou}, {Hauser}, {Haywood}, {Heiter}, {Helmi}, {Heu},
  {Hilger}, {Hobbs}, {Hofmann}, {Holland}, {Huckle}, {Hypki}, {Icardi},
  {Jan{\ss}en}, {Jevardat de Fombelle}, {Jonker}, {Juh{\'a}sz}, {Julbe},
  {Karampelas}, {Kewley}, {Klar}, {Kochoska}, {Kohley}, {Kolenberg},
  {Kontizas}, {Kontizas}, {Koposov}, {Kordopatis}, {Kostrzewa-Rutkowska},
  {Koubsky}, {Lambert}, {Lanza}, {Lasne}, {Lavigne}, {Le Fustec}, {Le
  Poncin-Lafitte}, {Lebreton}, {Leccia}, {Leclerc}, {Lecoeur-Taibi},
  {Lenhardt}, {Leroux}, {Liao}, {Licata}, {Lindstr{\o}m}, {Lister}, {Livanou},
  {Lobel}, {L{\'o}pez}, {Managau}, {Mann}, {Mantelet}, {Marchal}, {Marchant},
  {Marconi}, {Marinoni}, {Marschalk{\'o}}, {Marshall}, {Martino}, {Marton},
  {Mary}, {Massari}, {Matijevi{\v{c}}}, {Mazeh}, {McMillan}, {Messina},
  {Michalik}, {Millar}, {Molina}, {Molinaro}, {Moln{\'a}r}, {Montegriffo},
  {Mor}, {Morbidelli}, {Morel}, {Morris}, {Mulone}, {Muraveva}, {Musella},
  {Nelemans}, {Nicastro}, {Noval}, {O'Mullane}, {Ord{\'e}novic},
  {Ord{\'o}{\~n}ez-Blanco}, {Osborne}, {Pagani}, {Pagano}, {Pailler},
  {Palacin}, {Palaversa}, {Panahi}, {Pawlak}, {Piersimoni}, {Pineau}, {Plachy},
  {Plum}, {Poggio}, {Poujoulet}, {Pr{\v{s}}a}, {Pulone}, {Racero}, {Ragaini},
  {Rambaux}, {Ramos-Lerate}, {Regibo}, {Reyl{\'e}}, {Riclet}, {Ripepi}, {Riva},
  {Rivard}, {Rixon}, {Roegiers}, {Roelens}, {Romero-G{\'o}mez}, {Rowell},
  {Royer}, {Ruiz-Dern}, {Sadowski}, {Sagrist{\`a} Sell{\'e}s}, {Sahlmann},
  {Salgado}, {Salguero}, {Sanna}, {Santana-Ros}, {Sarasso}, {Savietto},
  {Schultheis}, {Sciacca}, {Segol}, {Segovia}, {S{\'e}gransan}, {Shih},
  {Siltala}, {Silva}, {Smart}, {Smith}, {Solano}, {Solitro}, {Sordo}, {Soria
  Nieto}, {Souchay}, {Spagna}, {Spoto}, {Stampa}, {Steele},
  {Steidelm{\"u}ller}, {Stephenson}, {Stoev}, {Suess}, {Surdej}, {Szabados},
  {Szegedi-Elek}, {Tapiador}, {Taris}, {Tauran}, {Taylor}, {Teixeira},
  {Terrett}, {Teyssand ier}, {Thuillot}, {Titarenko}, {Torra Clotet}, {Turon},
  {Ulla}, {Utrilla}, {Uzzi}, {Vaillant}, {Valentini}, {Valette}, {van Elteren},
  {Van Hemelryck}, {van Leeuwen}, {Vaschetto}, {Vecchiato}, {Veljanoski},
  {Viala}, {Vicente}, {Vogt}, {von Essen}, {Voss}, {Votruba}, {Voutsinas},
  {Walmsley}, {Weiler}, {Wertz}, {Wevers}, {Wyrzykowski}, {Yoldas},
  {{\v{Z}}erjal}, {Ziaeepour}, {Zorec}, {Zschocke}, {Zucker}, {Zurbach}, \&
  {Zwitter}}]{Gaia2018}
{\textit{Gaia} Collaboration}, {Brown}, A.~G.~A., {Vallenari}, A., {et~al.}
  2018,
  \href{http://dx.doi.org/10.1051/0004-6361/201833051}{\color{magenta}\aap},
  \href{https://ui.adsabs.harvard.edu/abs/2018A&A...616A...1G}{\color{blue}616},
  \href{https://ui.adsabs.harvard.edu/abs/2018A&A...616A...1G}{\color{blue}A1}

\bibitem[{{Gao} {et~al.}(2018){Gao}, {Marley}, \& {Ackerman}}]{gao2018}
{Gao}, P., {Marley}, M.~S., \& {Ackerman}, A.~S. 2018,
  \href{http://dx.doi.org/10.3847/1538-4357/aab0a1}{\color{magenta}\apj},
  \href{https://ui.adsabs.harvard.edu/abs/2018ApJ...855...86G}{\color{blue}855},
  \href{https://ui.adsabs.harvard.edu/abs/2018ApJ...855...86G}{\color{blue}86}

\bibitem[{{Ginzburg} \& {Chiang}(2019)}]{GinzburgChiang2019}
{Ginzburg}, S., \& {Chiang}, E. 2019,
  \href{http://dx.doi.org/10.1093/mnras/stz2901}{\color{magenta}\mnras},
  \href{https://ui.adsabs.harvard.edu/abs/2019MNRAS.490.4334G}{\color{blue}490},
  \href{https://ui.adsabs.harvard.edu/abs/2019MNRAS.490.4334G}{\color{blue}4334}

\bibitem[{{Greco} \& {Brandt}(2016)}]{Greco2016}
{Greco}, J.~P., \& {Brandt}, T.~D. 2016,
  \href{http://dx.doi.org/10.3847/1538-4357/833/2/134}{\color{magenta}\apj},
  \href{https://ui.adsabs.harvard.edu/abs/2016ApJ...833..134G}{\color{blue}833},
  \href{https://ui.adsabs.harvard.edu/abs/2016ApJ...833..134G}{\color{blue}134}

\bibitem[{{Haffert} {et~al.}(2019){Haffert}, {Bohn}, {de Boer}, {Snellen},
  {Brinchmann}, {Girard}, {Keller}, \& {Bacon}}]{Haffert2019}
{Haffert}, S.~Y., {Bohn}, A.~J., {de Boer}, J., {et~al.} 2019,
  \href{http://dx.doi.org/10.1038/s41550-019-0780-5}{\color{magenta}NatAs},
  \href{https://ui.adsabs.harvard.edu/abs/2019NatAs...3..749H}{\color{blue}3},
  \href{https://ui.adsabs.harvard.edu/abs/2019NatAs...3..749H}{\color{blue}749}

\bibitem[{{Hashimoto} {et~al.}(2020){Hashimoto}, {Aoyama}, {Konishi}, {Uyama},
  {Takasao}, {Ikoma}, \& {Tanigawa}}]{Hashimoto2020}
{Hashimoto}, J., {Aoyama}, Y., {Konishi}, M., {et~al.} 2020,
  \href{https://arxiv.org/abs/2003.07922}{\color{magenta}arXiv},
  \href{https://ui.adsabs.harvard.edu/abs/2020arXiv200307922H}{\color{blue}arXiv:2003.07922}

\bibitem[{{Hashimoto} {et~al.}(2012){Hashimoto}, {Dong}, {Kudo}, {Honda},
  {McClure}, {Zhu}, {Muto}, {Wisniewski}, {Abe}, {Brandner}, {Brandt},
  {Carson}, {Egner}, {Feldt}, {Fukagawa}, {Goto}, {Grady}, {Guyon}, {Hayano},
  {Hayashi}, {Hayashi}, {Henning}, {Hodapp}, {Ishii}, {Iye}, {Janson},
  {Kandori}, {Knapp}, {Kusakabe}, {Kuzuhara}, {Kwon}, {Matsuo}, {Mayama},
  {McElwain}, {Miyama}, {Morino}, {Moro-Martin}, {Nishimura}, {Pyo}, {Serabyn},
  {Suenaga}, {Suto}, {Suzuki}, {Takahashi}, {Takami}, {Takato}, {Terada},
  {Thalmann}, {Tomono}, {Turner}, {Watanabe}, {Yamada}, {Takami}, {Usuda}, \&
  {Tamura}}]{Hashimoto2012}
{Hashimoto}, J., {Dong}, R., {Kudo}, T., {et~al.} 2012,
  \href{http://dx.doi.org/10.1088/2041-8205/758/1/L19}{\color{magenta}\apjl},
  \href{https://ui.adsabs.harvard.edu/abs/2012ApJ...758L..19H}{\color{blue}758},
  \href{https://ui.adsabs.harvard.edu/abs/2012ApJ...758L..19H}{\color{blue}L19}

\bibitem[{{Hashimoto} {et~al.}(2015){Hashimoto}, {Tsukagoshi}, {Brown}, {Dong},
  {Muto}, {Zhu}, {Wisniewski}, {Ohashi}, {kudo}, {Kusakabe}, {Abe}, {Akiyama},
  {Brand ner}, {Brandt}, {Carson}, {Currie}, {Egner}, {Feldt}, {Grady},
  {Guyon}, {Hayano}, {Hayashi}, {Hayashi}, {Henning}, {Hodapp}, {Ishii}, {Iye},
  {Janson}, {Kand ori}, {Knapp}, {Kuzuhara}, {Kwon}, {Matsuo}, {McElwain},
  {Mayama}, {Mede}, {Miyama}, {Morino}, {Moro-Martin}, {Nishimura}, {Pyo},
  {Serabyn}, {Suenaga}, {Suto}, {Suzuki}, {Takahashi}, {Takami}, {Takato},
  {Terada}, {Thalmann}, {Tomono}, {Turner}, {Watanabe}, {Yamada}, {Takami},
  {Usuda}, \& {Tamura}}]{Hashimoto2015}
{Hashimoto}, J., {Tsukagoshi}, T., {Brown}, J.~M., {et~al.} 2015,
  \href{http://dx.doi.org/10.1088/0004-637X/799/1/43}{\color{magenta}\apj},
  \href{https://ui.adsabs.harvard.edu/abs/2015ApJ...799...43H}{\color{blue}799},
  \href{https://ui.adsabs.harvard.edu/abs/2015ApJ...799...43H}{\color{blue}43}

\bibitem[{{Helling} {et~al.}(2008){Helling}, {Dehn}, {Woitke}, \&
  {Hauschildt}}]{Helling2008}
{Helling}, C., {Dehn}, M., {Woitke}, P., \& {Hauschildt}, P.~H. 2008,
  \href{http://dx.doi.org/10.1086/533462}{\color{magenta}\apjl},
  \href{https://ui.adsabs.harvard.edu/abs/2008ApJ...675L.105H}{\color{blue}675},
  \href{https://ui.adsabs.harvard.edu/abs/2008ApJ...675L.105H}{\color{blue}L105}

\bibitem[{{Helling} \& {Woitke}(2006)}]{Helling2006}
{Helling}, C., \& {Woitke}, P. 2006,
  \href{http://dx.doi.org/10.1051/0004-6361:20054598}{\color{magenta}\aap},
  \href{https://ui.adsabs.harvard.edu/abs/2006A&A...455..325H}{\color{blue}455},
  \href{https://ui.adsabs.harvard.edu/abs/2006A&A...455..325H}{\color{blue}325}

\bibitem[{{Henden} {et~al.}(2015){Henden}, {Levine}, {Terrell}, \&
  {Welch}}]{Henden:vg}
{Henden}, A.~A., {Levine}, S., {Terrell}, D., \& {Welch}, D.~L. 2015, AAS
  Meeting,
  \href{https://ui.adsabs.harvard.edu/abs/2015AAS...22533616H}{\color{blue}225},
  \href{https://ui.adsabs.harvard.edu/abs/2015AAS...22533616H}{\color{blue}336.16}

\bibitem[{{Huby} {et~al.}(2017){Huby}, {Bottom}, {Femenia}, {Ngo}, {Mawet},
  {Serabyn}, \& {Absil}}]{Huby2017}
{Huby}, E., {Bottom}, M., {Femenia}, B., {et~al.} 2017,
  \href{http://dx.doi.org/10.1051/0004-6361/201630232}{\color{magenta}\aap},
  \href{http://adsabs.harvard.edu/abs/2017A%26A...600A..46H}{\color{blue}600},
  \href{http://adsabs.harvard.edu/abs/2017A%26A...600A..46H}{\color{blue}A46}

\bibitem[{{Isella} {et~al.}(2019){Isella}, {Benisty}, {Teague}, {Bae},
  {Keppler}, {Facchini}, \& {P{\'e}rez}}]{Isella2019}
{Isella}, A., {Benisty}, M., {Teague}, R., {et~al.} 2019,
  \href{http://dx.doi.org/10.3847/2041-8213/ab2a12}{\color{magenta}\apjl},
  \href{https://ui.adsabs.harvard.edu/abs/2019ApJ...879L..25I}{\color{blue}879},
  \href{https://ui.adsabs.harvard.edu/abs/2019ApJ...879L..25I}{\color{blue}L25}

\bibitem[{{Keppler} {et~al.}(2018){Keppler}, {Benisty}, {M{\"u}ller},
  {Henning}, {van Boekel}, {Cantalloube}, {Ginski}, {van Holstein}, {Maire},
  {Pohl}, {Samland }, {Avenhaus}, {Baudino}, {Boccaletti}, {de Boer},
  {Bonnefoy}, {Chauvin}, {Desidera}, {Langlois}, {Lazzoni}, {Marleau},
  {Mordasini}, {Pawellek}, {Stolker}, {Vigan}, {Zurlo}, {Birnstiel},
  {Brandner}, {Feldt}, {Flock}, {Girard}, {Gratton}, {Hagelberg}, {Isella},
  {Janson}, {Juhasz}, {Kemmer}, {Kral}, {Lagrange}, {Launhardt}, {Matter},
  {M{\'e}nard}, {Milli}, {Molli{\`e}re}, {Olofsson}, {P{\'e}rez}, {Pinilla},
  {Pinte}, {Quanz}, {Schmidt}, {Udry}, {Wahhaj}, {Williams}, {Buenzli},
  {Cudel}, {Dominik}, {Galicher}, {Kasper}, {Lannier}, {Mesa}, {Mouillet},
  {Peretti}, {Perrot}, {Salter}, {Sissa}, {Wildi}, {Abe}, {Antichi},
  {Augereau}, {Baruffolo}, {Baudoz}, {Bazzon}, {Beuzit}, {Blanchard}, {Brems},
  {Buey}, {De Caprio}, {Carbillet}, {Carle}, {Cascone}, {Cheetham}, {Claudi},
  {Costille}, {Delboulb{\'e}}, {Dohlen}, {Fantinel}, {Feautrier}, {Fusco},
  {Giro}, {Gluck}, {Gry}, {Hubin}, {Hugot}, {Jaquet}, {Le Mignant}, {Llored},
  {Madec}, {Magnard}, {Martinez}, {Maurel}, {Meyer}, {M{\"o}ller-Nilsson},
  {Moulin}, {Mugnier}, {Orign{\'e}}, {Pavlov}, {Perret}, {Petit}, {Pragt},
  {Puget}, {Rabou}, {Ramos}, {Rigal}, {Rochat}, {Roelfsema}, {Rousset}, {Roux},
  {Salasnich}, {Sauvage}, {Sevin}, {Soenke}, {Stadler}, {Suarez}, {Turatto}, \&
  {Weber}}]{Keppler2018}
{Keppler}, M., {Benisty}, M., {M{\"u}ller}, A., {et~al.} 2018,
  \href{http://dx.doi.org/10.1051/0004-6361/201832957}{\color{magenta}\aap},
  \href{https://ui.adsabs.harvard.edu/abs/2018A&A...617A..44K}{\color{blue}617},
  \href{https://ui.adsabs.harvard.edu/abs/2018A&A...617A..44K}{\color{blue}A44}

\bibitem[{{Keppler} {et~al.}(2019){Keppler}, {Teague}, {Bae}, {Benisty},
  {Henning}, {van Boekel}, {Chapillon}, {Pinilla}, {Williams}, {Bertrang},
  {Facchini}, {Flock}, {Ginski}, {Juhasz}, {Klahr}, {Liu}, {M{\"u}ller},
  {P{\'e}rez}, {Pohl}, {Rosotti}, {Samland}, \& {Semenov}}]{Keppler2019}
{Keppler}, M., {Teague}, R., {Bae}, J., {et~al.} 2019,
  \href{http://dx.doi.org/10.1051/0004-6361/201935034}{\color{magenta}\aap},
  \href{https://ui.adsabs.harvard.edu/abs/2019A&A...625A.118K}{\color{blue}625},
  \href{https://ui.adsabs.harvard.edu/abs/2019A&A...625A.118K}{\color{blue}A118}

\bibitem[{{Kraus} \& {Ireland}(2012)}]{Kraus2012}
{Kraus}, A.~L., \& {Ireland}, M.~J. 2012,
  \href{http://dx.doi.org/10.1088/0004-637X/745/1/5}{\color{magenta}\apj},
  \href{https://ui.adsabs.harvard.edu/abs/2012ApJ...745....5K}{\color{blue}745},
  \href{https://ui.adsabs.harvard.edu/abs/2012ApJ...745....5K}{\color{blue}5}

\bibitem[{{Li} \& {Greenberg}(1998)}]{li98}
{Li}, A., \& {Greenberg}, J.~M. 1998, \aap,
  \href{http://ads.bao.ac.cn/abs/1998A%26A...331..291L}{\color{blue}331},
  \href{http://ads.bao.ac.cn/abs/1998A%26A...331..291L}{\color{blue}291}

\bibitem[{{Liu}(2004)}]{Liu2004}
{Liu}, M.~C. 2004,
  \href{http://dx.doi.org/10.1126/science.1102929}{\color{magenta}Science},
  \href{https://ui.adsabs.harvard.edu/abs/2004Sci...305.1442L}{\color{blue}305},
  \href{https://ui.adsabs.harvard.edu/abs/2004Sci...305.1442L}{\color{blue}1442}

\bibitem[{{Liu} {et~al.}(2016){Liu}, {Dupuy}, \& {Allers}}]{liu2016}
{Liu}, M.~C., {Dupuy}, T.~J., \& {Allers}, K.~N. 2016,
  \href{http://dx.doi.org/10.3847/1538-4357/833/1/96}{\color{magenta}\apj},
  \href{https://ui.adsabs.harvard.edu/abs/2016ApJ...833...96L}{\color{blue}833},
  \href{https://ui.adsabs.harvard.edu/abs/2016ApJ...833...96L}{\color{blue}96}

\bibitem[{{Liu} {et~al.}(2013){Liu}, {Magnier}, {Deacon}, {Allers}, {Dupuy},
  {Kotson}, {Aller}, {Burgett}, {Chambers}, {Draper}, {Hodapp}, {Jedicke},
  {Kaiser}, {Kudritzki}, {Metcalfe}, {Morgan}, {Price}, {Tonry}, \&
  {Wainscoat}}]{Liu2013}
{Liu}, M.~C., {Magnier}, E.~A., {Deacon}, N.~R., {et~al.} 2013,
  \href{http://dx.doi.org/10.1088/2041-8205/777/2/L20}{\color{magenta}\apjl},
  \href{https://ui.adsabs.harvard.edu/abs/2013ApJ...777L..20L}{\color{blue}777},
  \href{https://ui.adsabs.harvard.edu/abs/2013ApJ...777L..20L}{\color{blue}L20}

\bibitem[{{Marley} {et~al.}(2007){Marley}, {Fortney}, {Hubickyj},
  {Bodenheimer}, \& {Lissauer}}]{Marley2007}
{Marley}, M.~S., {Fortney}, J.~J., {Hubickyj}, O., {et~al.} 2007,
  \href{http://dx.doi.org/10.1086/509759}{\color{magenta}\apj},
  \href{https://ui.adsabs.harvard.edu/abs/2007ApJ...655..541M}{\color{blue}655},
  \href{https://ui.adsabs.harvard.edu/abs/2007ApJ...655..541M}{\color{blue}541}

\bibitem[{{Marley} {et~al.}(2012){Marley}, {Saumon}, {Cushing}, {Ackerman},
  {Fortney}, \& {Freedman}}]{Marley2012}
{Marley}, M.~S., {Saumon}, D., {Cushing}, M., {et~al.} 2012,
  \href{http://dx.doi.org/10.1088/0004-637X/754/2/135}{\color{magenta}\apj},
  \href{https://ui.adsabs.harvard.edu/abs/2012ApJ...754..135M}{\color{blue}754},
  \href{https://ui.adsabs.harvard.edu/abs/2012ApJ...754..135M}{\color{blue}135}

\bibitem[{{Marley} {et~al.}(2010){Marley}, {Saumon}, \&
  {Goldblatt}}]{marley2010}
{Marley}, M.~S., {Saumon}, D., \& {Goldblatt}, C. 2010,
  \href{http://dx.doi.org/10.1088/2041-8205/723/1/L117}{\color{magenta}\apjl},
  \href{https://ui.adsabs.harvard.edu/abs/2010ApJ...723L.117M}{\color{blue}723},
  \href{https://ui.adsabs.harvard.edu/abs/2010ApJ...723L.117M}{\color{blue}L117}

\bibitem[{{Marocco} {et~al.}(2014){Marocco}, {Day-Jones}, {Lucas}, {Jones},
  {Smart}, {Zhang}, {Gomes}, {Burningham}, {Pinfield}, {Raddi}, \&
  {Smith}}]{Marocco2014}
{Marocco}, F., {Day-Jones}, A.~C., {Lucas}, P.~W., {et~al.} 2014,
  \href{http://dx.doi.org/10.1093/mnras/stt2463}{\color{magenta}\mnras},
  \href{https://ui.adsabs.harvard.edu/abs/2014MNRAS.439..372M}{\color{blue}439},
  \href{https://ui.adsabs.harvard.edu/abs/2014MNRAS.439..372M}{\color{blue}372}

\bibitem[{{Marois} {et~al.}(2006){Marois}, {Lafreni{\`e}re}, {Doyon},
  {Macintosh}, \& {Nadeau}}]{Marois2006}
{Marois}, C., {Lafreni{\`e}re}, D., {Doyon}, R., {et~al.} 2006,
  \href{http://dx.doi.org/10.1086/500401}{\color{magenta}\apj},
  \href{https://ui.adsabs.harvard.edu/abs/2006ApJ...641..556M}{\color{blue}641},
  \href{https://ui.adsabs.harvard.edu/abs/2006ApJ...641..556M}{\color{blue}556}

\bibitem[{{Mathis}(1990)}]{Mathis1990}
{Mathis}, J.~S. 1990,
  \href{http://dx.doi.org/10.1146/annurev.aa.28.090190.000345}{\color{magenta}\araa},
  \href{https://ui.adsabs.harvard.edu/abs/1990ARA&A..28...37M}{\color{blue}28},
  \href{https://ui.adsabs.harvard.edu/abs/1990ARA&A..28...37M}{\color{blue}37}

\bibitem[{{Mendigut{\'\i}a} {et~al.}(2018){Mendigut{\'\i}a}, {Oudmaijer},
  {Schneider}, {Hu{\'e}lamo}, {Baines}, {Brittain}, \&
  {Aberasturi}}]{Mendigutia2018}
{Mendigut{\'\i}a}, I., {Oudmaijer}, R.~D., {Schneider}, P.~C., {et~al.} 2018,
  \href{http://dx.doi.org/10.1051/0004-6361/201834233}{\color{magenta}\aap},
  \href{https://ui.adsabs.harvard.edu/abs/2018A&A...618L...9M}{\color{blue}618},
  \href{https://ui.adsabs.harvard.edu/abs/2018A&A...618L...9M}{\color{blue}L9}

\bibitem[{{Mesa} {et~al.}(2019){Mesa}, {Keppler}, {Cantalloube}, {Rodet},
  {Charnay}, {Gratton}, {Langlois}, {Boccaletti}, {Bonnefoy}, {Vigan},
  {Flasseur}, {Bae}, {Benisty}, {Chauvin}, {de Boer}, {Desidera}, {Henning},
  {Lagrange}, {Meyer}, {Milli}, {M{\"u}ller}, {Pairet}, {Zurlo}, {Antoniucci},
  {Baudino}, {Brown Sevilla}, {Cascone}, {Cheetham}, {Claudi}, {Delorme},
  {D'Orazi}, {Feldt}, {Hagelberg}, {Janson}, {Kral}, {Lagadec}, {Lazzoni},
  {Ligi}, {Maire}, {Martinez}, {Menard}, {Meunier}, {Perrot}, {Petrus},
  {Pinte}, {Rickman}, {Rochat}, {Rouan}, {Samland}, {Sauvage}, {Schmidt},
  {Udry}, {Weber}, \& {Wildi}}]{Mesa2019}
{Mesa}, D., {Keppler}, M., {Cantalloube}, F., {et~al.} 2019,
  \href{http://dx.doi.org/10.1051/0004-6361/201936764}{\color{magenta}\aap},
  \href{https://ui.adsabs.harvard.edu/abs/2019A&A...632A..25M}{\color{blue}632},
  \href{https://ui.adsabs.harvard.edu/abs/2019A&A...632A..25M}{\color{blue}A25}

\bibitem[{{Mie}(1908)}]{mie1908}
{Mie}, G. 1908,
  \href{http://dx.doi.org/10.1002/andp.19083300302}{\color{magenta}AnP},
  \href{http://ads.bao.ac.cn/abs/1908AnP...330..377M}{\color{blue}330},
  \href{http://ads.bao.ac.cn/abs/1908AnP...330..377M}{\color{blue}377}

\bibitem[{{Mordasini}(2014)}]{Mordasini2014}
{Mordasini}, C. 2014,
  \href{http://dx.doi.org/10.1051/0004-6361/201423702}{\color{magenta}\aap},
  \href{https://ui.adsabs.harvard.edu/abs/2014A%26A...572A.118M}{\color{blue}572},
  \href{https://ui.adsabs.harvard.edu/abs/2014A%26A...572A.118M}{\color{blue}A118}

\bibitem[{{Mordasini} {et~al.}(2017){Mordasini}, {Marleau}, \&
  {Molli{\`e}re}}]{Mordasini2017}
{Mordasini}, C., {Marleau}, G.~D., \& {Molli{\`e}re}, P. 2017,
  \href{http://dx.doi.org/10.1051/0004-6361/201630077}{\color{magenta}\aap},
  \href{https://ui.adsabs.harvard.edu/abs/2017A&A...608A..72M}{\color{blue}608},
  \href{https://ui.adsabs.harvard.edu/abs/2017A&A...608A..72M}{\color{blue}A72}

\bibitem[{{Moses} {et~al.}(2016){Moses}, {Marley}, {Zahnle}, {Line}, {Fortney},
  {Barman}, {Visscher}, {Lewis}, \& {Wolff}}]{moses2016}
{Moses}, J.~I., {Marley}, M.~S., {Zahnle}, K., {et~al.} 2016,
  \href{http://dx.doi.org/10.3847/0004-637X/829/2/66}{\color{magenta}\apj},
  \href{https://ui.adsabs.harvard.edu/abs/2016ApJ...829...66M}{\color{blue}829},
  \href{https://ui.adsabs.harvard.edu/abs/2016ApJ...829...66M}{\color{blue}66}

\bibitem[{{Movshovitz} {et~al.}(2010){Movshovitz}, {Bodenheimer}, {Podolak}, \&
  {Lissauer}}]{Movshovitz2010}
{Movshovitz}, N., {Bodenheimer}, P., {Podolak}, M., \& {Lissauer}, J.~J. 2010,
  \href{http://dx.doi.org/10.1016/j.icarus.2010.06.009}{\color{magenta}\icarus},
  \href{https://ui.adsabs.harvard.edu/abs/2010Icar..209..616M}{\color{blue}209},
  \href{https://ui.adsabs.harvard.edu/abs/2010Icar..209..616M}{\color{blue}616}

\bibitem[{{M{\"u}ller} {et~al.}(2018){M{\"u}ller}, {Keppler}, {Henning},
  {Samland}, {Chauvin}, {Beust}, {Maire}, {Molaverdikhani}, {van Boekel},
  {Benisty}, {Boccaletti}, {Bonnefoy}, {Cantalloube}, {Charnay}, {Baudino},
  {Gennaro}, {Long}, {Cheetham}, {Desidera}, {Feldt}, {Fusco}, {Girard},
  {Gratton}, {Hagelberg}, {Janson}, {Lagrange}, {Langlois}, {Lazzoni}, {Ligi},
  {M{\'e}nard}, {Mesa}, {Meyer}, {Molli{\`e}re}, {Mordasini}, {Moulin},
  {Pavlov}, {Pawellek}, {Quanz}, {Ramos}, {Rouan}, {Sissa}, {Stadler}, {Vigan},
  {Wahhaj}, {Weber}, \& {Zurlo}}]{Muller2018}
{M{\"u}ller}, A., {Keppler}, M., {Henning}, T., {et~al.} 2018,
  \href{http://dx.doi.org/10.1051/0004-6361/201833584}{\color{magenta}\aap},
  \href{https://ui.adsabs.harvard.edu/abs/2018A&A...617L...2M}{\color{blue}617},
  \href{https://ui.adsabs.harvard.edu/abs/2018A&A...617L...2M}{\color{blue}L2}

\bibitem[{{Natta} {et~al.}(2004){Natta}, {Testi}, {Muzerolle}, {Randich},
  {Comer{\'o}n}, \& {Persi}}]{Natta2004}
{Natta}, A., {Testi}, L., {Muzerolle}, J., {et~al.} 2004,
  \href{http://dx.doi.org/10.1051/0004-6361:20040356}{\color{magenta}\aap},
  \href{https://ui.adsabs.harvard.edu/abs/2004A&A...424..603N}{\color{blue}424},
  \href{https://ui.adsabs.harvard.edu/abs/2004A&A...424..603N}{\color{blue}603}

\bibitem[{{Nielsen} {et~al.}(2017){Nielsen}, {De Rosa}, {Rameau}, {Wang},
  {Esposito}, {Millar-Blanchaer}, {Marois}, {Vigan}, {Ammons}, {Artigau},
  {Bailey}, {Blunt}, {Bulger}, {Chilcote}, {Cotten}, {Doyon}, {Duch{\^e}ne},
  {Fabrycky}, {Fitzgerald}, {Follette}, {Gerard}, {Goodsell}, {Graham},
  {Greenbaum}, {Hibon}, {Hinkley}, {Hung}, {Ingraham}, {Jensen-Clem}, {Kalas},
  {Konopacky}, {Larkin}, {Macintosh}, {Maire}, {Marchis}, {Metchev},
  {Morzinski}, {Murray-Clay}, {Oppenheimer}, {Palmer}, {Patience}, {Perrin},
  {Poyneer}, {Pueyo}, {Rafikov}, {Rajan}, {Rantakyr{\"o}}, {Ruffio},
  {Savransky}, {Schneider}, {Sivaramakrishnan}, {Song}, {Soummer}, {Thomas},
  {Wallace}, {Ward-Duong}, {Wiktorowicz}, \& {Wolff}}]{Nielsen:2017kf}
{Nielsen}, E.~L., {De Rosa}, R.~J., {Rameau}, J., {et~al.} 2017,
  \href{http://dx.doi.org/10.3847/1538-3881/aa8a69}{\color{magenta}\aj},
  \href{https://ui.adsabs.harvard.edu/abs/2017AJ....154..218N}{\color{blue}154},
  \href{https://ui.adsabs.harvard.edu/abs/2017AJ....154..218N}{\color{blue}218}

\bibitem[{{Ormel}(2014)}]{Ormel2014}
{Ormel}, C.~W. 2014,
  \href{http://dx.doi.org/10.1088/2041-8205/789/1/L18}{\color{magenta}\apjl},
  \href{https://ui.adsabs.harvard.edu/abs/2014ApJ...789L..18O}{\color{blue}789},
  \href{https://ui.adsabs.harvard.edu/abs/2014ApJ...789L..18O}{\color{blue}L18}

\bibitem[{{Pecaut} \& {Mamajek}(2016)}]{Pecaut2016}
{Pecaut}, M.~J., \& {Mamajek}, E.~E. 2016, VizieR Online Data Catalog,
  \href{https://ui.adsabs.harvard.edu/abs/2016yCat..74610794P}{\color{blue}J/MNRAS/461/794}

\bibitem[{{Pinte} {et~al.}(2009){Pinte}, {Harries}, {Min}, {Watson},
  {Dullemond}, {Woitke}, {M{\'e}nard}, \& {Dur{\'a}n-Rojas}}]{Pinte2009}
{Pinte}, C., {Harries}, T.~J., {Min}, M., {et~al.} 2009,
  \href{http://dx.doi.org/10.1051/0004-6361/200811555}{\color{magenta}\aap},
  \href{https://ui.adsabs.harvard.edu/abs/2009A&A...498..967P}{\color{blue}498},
  \href{https://ui.adsabs.harvard.edu/abs/2009A&A...498..967P}{\color{blue}967}

\bibitem[{{Pinte} {et~al.}(2006){Pinte}, {M{\'e}nard}, {Duch{\^e}ne}, \&
  {Bastien}}]{Pinte2006}
{Pinte}, C., {M{\'e}nard}, F., {Duch{\^e}ne}, G., \& {Bastien}, P. 2006,
  \href{http://dx.doi.org/10.1051/0004-6361:20053275}{\color{magenta}\aap},
  \href{https://ui.adsabs.harvard.edu/abs/2006A&A...459..797P}{\color{blue}459},
  \href{https://ui.adsabs.harvard.edu/abs/2006A&A...459..797P}{\color{blue}797}

\bibitem[{{Piso} \& {Youdin}(2014)}]{Piso2014}
{Piso}, A.-M.~A., \& {Youdin}, A.~N. 2014,
  \href{http://dx.doi.org/10.1088/0004-637X/786/1/21}{\color{magenta}\apj},
  \href{https://ui.adsabs.harvard.edu/abs/2014ApJ...786...21P}{\color{blue}786},
  \href{https://ui.adsabs.harvard.edu/abs/2014ApJ...786...21P}{\color{blue}21}

\bibitem[{{Piso} {et~al.}(2015){Piso}, {Youdin}, \& {Murray-Clay}}]{Piso2015}
{Piso}, A.-M.~A., {Youdin}, A.~N., \& {Murray-Clay}, R.~A. 2015,
  \href{http://dx.doi.org/10.1088/0004-637X/800/2/82}{\color{magenta}\apj},
  \href{https://ui.adsabs.harvard.edu/abs/2015ApJ...800...82P}{\color{blue}800},
  \href{https://ui.adsabs.harvard.edu/abs/2015ApJ...800...82P}{\color{blue}82}

\bibitem[{{Pollack} {et~al.}(1996){Pollack}, {Hubickyj}, {Bodenheimer},
  {Lissauer}, {Podolak}, \& {Greenzweig}}]{Pollack1996}
{Pollack}, J.~B., {Hubickyj}, O., {Bodenheimer}, P., {et~al.} 1996,
  \href{http://dx.doi.org/10.1006/icar.1996.0190}{\color{magenta}\icarus},
  \href{https://ui.adsabs.harvard.edu/abs/1996Icar..124...62P}{\color{blue}124},
  \href{https://ui.adsabs.harvard.edu/abs/1996Icar..124...62P}{\color{blue}62}

\bibitem[{{Pueyo}(2016)}]{Pueyo2016}
{Pueyo}, L. 2016,
  \href{http://dx.doi.org/10.3847/0004-637X/824/2/117}{\color{magenta}\apj},
  \href{http://ui.adsabs.harvard.edu/abs/2016ApJ...824..117P}{\color{blue}824},
  \href{http://ui.adsabs.harvard.edu/abs/2016ApJ...824..117P}{\color{blue}117}

\bibitem[{{Quanz} {et~al.}(2013){Quanz}, {Amara}, {Meyer}, {Kenworthy},
  {Kasper}, \& {Girard}}]{Quanz2013}
{Quanz}, S.~P., {Amara}, A., {Meyer}, M.~R., {et~al.} 2013,
  \href{http://dx.doi.org/10.1088/2041-8205/766/1/L1}{\color{magenta}\apjl},
  \href{https://ui.adsabs.harvard.edu/abs/2013ApJ...766L...1Q}{\color{blue}766},
  \href{https://ui.adsabs.harvard.edu/abs/2013ApJ...766L...1Q}{\color{blue}L1}

\bibitem[{{Rajan} {et~al.}(2017){Rajan}, {Rameau}, {De Rosa}, {Marley},
  {Graham}, {Macintosh}, {Marois}, {Morley}, {Patience}, {Pueyo}, {Saumon},
  {Ward-Duong}, {Ammons}, {Arriaga}, {Bailey}, {Barman}, {Bulger}, {Burrows},
  {Chilcote}, {Cotten}, {Czekala}, {Doyon}, {Duch{\^e}ne}, {Esposito},
  {Fitzgerald}, {Follette}, {Fortney}, {Goodsell}, {Greenbaum}, {Hibon},
  {Hung}, {Ingraham}, {Johnson-Groh}, {Kalas}, {Konopacky}, {Lafreni{\`e}re},
  {Larkin}, {Maire}, {Marchis}, {Metchev}, {Millar-Blanchaer}, {Morzinski},
  {Nielsen}, {Oppenheimer}, {Palmer}, {Patel}, {Perrin}, {Poyneer},
  {Rantakyr{\"o}}, {Ruffio}, {Savransky}, {Schneider}, {Sivaramakrishnan},
  {Song}, {Soummer}, {Thomas}, {Vasisht}, {Wallace}, {Wang}, {Wiktorowicz}, \&
  {Wolff}}]{rajan2017}
{Rajan}, A., {Rameau}, J., {De Rosa}, R.~J., {et~al.} 2017,
  \href{http://dx.doi.org/10.3847/1538-3881/aa74db}{\color{magenta}\aj},
  \href{https://ui.adsabs.harvard.edu/abs/2017AJ....154...10R}{\color{blue}154},
  \href{https://ui.adsabs.harvard.edu/abs/2017AJ....154...10R}{\color{blue}10}

\bibitem[{{Rameau} {et~al.}(2017){Rameau}, {Follette}, {Pueyo}, {Marois},
  {Macintosh}, {Millar-Blanchaer}, {Wang}, {Vega}, {Doyon}, {Lafreni{\`e}re},
  {Nielsen}, {Bailey}, {Chilcote}, {Close}, {Esposito}, {Males}, {Metchev},
  {Morzinski}, {Ruffio}, {Wolff}, {Ammons}, {Barman}, {Bulger}, {Cotten}, {De
  Rosa}, {Duchene}, {Fitzgerald}, {Goodsell}, {Graham}, {Greenbaum}, {Hibon},
  {Hung}, {Ingraham}, {Kalas}, {Konopacky}, {Larkin}, {Maire}, {Marchis},
  {Oppenheimer}, {Palmer}, {Patience}, {Perrin}, {Poyneer}, {Rajan},
  {Rantakyr{\"o}}, {Marley}, {Savransky}, {Schneider}, {Sivaramakrishnan},
  {Song}, {Soummer}, {Thomas}, {Wallace}, {Ward-Duong}, \&
  {Wiktorowicz}}]{Rameau2017}
{Rameau}, J., {Follette}, K.~B., {Pueyo}, L., {et~al.} 2017,
  \href{http://dx.doi.org/10.3847/1538-3881/aa6cae}{\color{magenta}\aj},
  \href{https://ui.adsabs.harvard.edu/abs/2017AJ....153..244R}{\color{blue}153},
  \href{https://ui.adsabs.harvard.edu/abs/2017AJ....153..244R}{\color{blue}244}

\bibitem[{{Ren} {et~al.}(2019){Ren}, {Choquet}, {Perrin}, {Duch{\^e}ne},
  {Debes}, {Pueyo}, {Rice}, {Chen}, {Schneider}, {Esposito}, {Poteet}, {Wang},
  {Ammons}, {Ansdell}, {Arriaga}, {Bailey}, {Barman}, {Sebasti{\'a}n Bruzzone},
  {Bulger}, {Chilcote}, {Cotten}, {De Rosa}, {Doyon}, {Fitzgerald}, {Follette},
  {Goodsell}, {Gerard}, {Graham}, {Greenbaum}, {Hagan}, {Hibon}, {Hines},
  {Hung}, {Ingraham}, {Kalas}, {Konopacky}, {Larkin}, {Macintosh}, {Maire},
  {Marchis}, {Marois}, {Mazoyer}, {M{\'e}nard}, {Metchev}, {Millar-Blanchaer},
  {Mittal}, {Moerchen}, {Nielsen}, {N{\textquoteright}Diaye}, {Oppenheimer},
  {Palmer}, {Patience}, {Pinte}, {Poyneer}, {Rajan}, {Rameau}, {Rantakyr{\"o}},
  {Ruffio}, {Ryan}, {Savransky}, {Schneider}, {Sivaramakrishnan}, {Song},
  {Soummer}, {Stark}, {Thomas}, {Vigan}, {Wallace}, {Ward-Duong},
  {Wiktorowicz}, {Wolff}, {Ygouf}, \& {Norman}}]{Ren2019}
{Ren}, B., {Choquet}, {\'E}., {Perrin}, M.~D., {et~al.} 2019,
  \href{http://dx.doi.org/10.3847/1538-4357/ab3403}{\color{magenta}\apj},
  \href{https://ui.adsabs.harvard.edu/abs/2019ApJ...882...64R}{\color{blue}882},
  \href{https://ui.adsabs.harvard.edu/abs/2019ApJ...882...64R}{\color{blue}64}

\bibitem[{{Rigliaco} {et~al.}(2012){Rigliaco}, {Natta}, {Testi}, {Randich},
  {Alcal{\`a}}, {Covino}, \& {Stelzer}}]{Rigliaco2012}
{Rigliaco}, E., {Natta}, A., {Testi}, L., {et~al.} 2012,
  \href{http://dx.doi.org/10.1051/0004-6361/201219832}{\color{magenta}\aap},
  \href{https://ui.adsabs.harvard.edu/abs/2012A%26A...548A..56R}{\color{blue}548},
  \href{https://ui.adsabs.harvard.edu/abs/2012A%26A...548A..56R}{\color{blue}A56}

\bibitem[{{Rouleau} \& {Martin}(1991)}]{rouleau91}
{Rouleau}, F., \& {Martin}, P.~G. 1991,
  \href{http://dx.doi.org/10.1086/170382}{\color{magenta}\apj},
  \href{http://ads.bao.ac.cn/abs/1991ApJ...377..526R}{\color{blue}377},
  \href{http://ads.bao.ac.cn/abs/1991ApJ...377..526R}{\color{blue}526}

\bibitem[{{Ruane} {et~al.}(2019){Ruane}, {Ngo}, {Mawet}, {Absil}, {Choquet},
  {Cook}, {Gomez Gonzalez}, {Huby}, {Matthews}, {Meshkat}, {Reggiani},
  {Serabyn}, {Wallack}, \& {Xuan}}]{Ruane2019}
{Ruane}, G., {Ngo}, H., {Mawet}, D., {et~al.} 2019,
  \href{http://dx.doi.org/10.3847/1538-3881/aafee2}{\color{magenta}\aj},
  \href{https://ui.adsabs.harvard.edu/abs/2019AJ....157..118R}{\color{blue}157},
  \href{https://ui.adsabs.harvard.edu/abs/2019AJ....157..118R}{\color{blue}118}

\bibitem[{{Sallum} {et~al.}(2015){Sallum}, {Follette}, {Eisner}, {Close},
  {Hinz}, {Kratter}, {Males}, {Skemer}, {Macintosh}, {Tuthill}, {Bailey},
  {Defr{\`e}re}, {Morzinski}, {Rodigas}, {Spalding}, {Vaz}, \&
  {Weinberger}}]{Sallum2015}
{Sallum}, S., {Follette}, K.~B., {Eisner}, J.~A., {et~al.} 2015,
  \href{http://dx.doi.org/10.1038/nature15761}{\color{magenta}\nat},
  \href{https://ui.adsabs.harvard.edu/abs/2015Natur.527..342S}{\color{blue}527},
  \href{https://ui.adsabs.harvard.edu/abs/2015Natur.527..342S}{\color{blue}342}

\bibitem[{{Samland} {et~al.}(2017){Samland}, {Molli{\`e}re}, {Bonnefoy},
  {Maire}, {Cantalloube}, {Cheetham}, {Mesa}, {Gratton}, {Biller}, {Wahhaj},
  {Bouwman}, {Brandner}, {Melnick}, {Carson}, {Janson}, {Henning}, {Homeier},
  {Mordasini}, {Langlois}, {Quanz}, {van Boekel}, {Zurlo}, {Schlieder},
  {Avenhaus}, {Beuzit}, {Boccaletti}, {Bonavita}, {Chauvin}, {Claudi}, {Cudel},
  {Desidera}, {Feldt}, {Fusco}, {Galicher}, {Kopytova}, {Lagrange}, {Le
  Coroller}, {Martinez}, {Moeller-Nilsson}, {Mouillet}, {Mugnier}, {Perrot},
  {Sevin}, {Sissa}, {Vigan}, \& {Weber}}]{Samland2017}
{Samland}, M., {Molli{\`e}re}, P., {Bonnefoy}, M., {et~al.} 2017,
  \href{http://dx.doi.org/10.1051/0004-6361/201629767}{\color{magenta}\aap},
  \href{https://ui.adsabs.harvard.edu/abs/2017A&A...603A..57S}{\color{blue}603},
  \href{https://ui.adsabs.harvard.edu/abs/2017A&A...603A..57S}{\color{blue}A57}

\bibitem[{{Serabyn} {et~al.}(2017){Serabyn}, {Huby}, {Matthews}, {Mawet},
  {Absil}, {Femenia}, {Wizinowich}, {Karlsson}, {Bottom}, {Campbell},
  {Carlomagno}, {Defr{\`e}re}, {Delacroix}, {Forsberg}, {Gomez Gonzalez},
  {Habraken}, {Jolivet}, {Liewer}, {Lilley}, {Piron}, {Reggiani}, {Surdej},
  {Tran}, {Vargas Catal{\'a}n}, \& {Wertz}}]{Serabyn2017}
{Serabyn}, E., {Huby}, E., {Matthews}, K., {et~al.} 2017,
  \href{http://dx.doi.org/10.3847/1538-3881/153/1/43}{\color{magenta}\aj},
  \href{http://adsabs.harvard.edu/abs/2017AJ....153...43S}{\color{blue}153},
  \href{http://adsabs.harvard.edu/abs/2017AJ....153...43S}{\color{blue}43}

\bibitem[{{Service} {et~al.}(2016){Service}, {Lu}, {Campbell}, {Sitarski},
  {Ghez}, \& {Anderson}}]{Service2016}
{Service}, M., {Lu}, J.~R., {Campbell}, R., {et~al.} 2016,
  \href{http://dx.doi.org/10.1088/1538-3873/128/967/095004}{\color{magenta}\pasp},
  \href{http://adsabs.harvard.edu/abs/2016PASP..128i5004S}{\color{blue}128},
  \href{http://adsabs.harvard.edu/abs/2016PASP..128i5004S}{\color{blue}095004}

\bibitem[{Skrutskie {et~al.}(2006)Skrutskie, Cutri, Stiening, Weinberg,
  Schneider, Carpenter, Beichman, Capps, Chester, Elias, Huchra, Liebert,
  Lonsdale, Monet, Price, \& Seitzer}]{Skrutskie:2006hla}
Skrutskie, M.~F., Cutri, R.~M., Stiening, R., {et~al.} 2006, AJ, 131, 1163

\bibitem[{{Soummer} {et~al.}(2012){Soummer}, {Pueyo}, \&
  {Larkin}}]{Soummer2012}
{Soummer}, R., {Pueyo}, L., \& {Larkin}, J. 2012,
  \href{http://dx.doi.org/10.1088/2041-8205/755/2/L28}{\color{magenta}\apjl},
  \href{https://ui.adsabs.harvard.edu/abs/2012ApJ...755L..28S}{\color{blue}755},
  \href{https://ui.adsabs.harvard.edu/abs/2012ApJ...755L..28S}{\color{blue}L28}

\bibitem[{{Stephens} {et~al.}(2009){Stephens}, {Leggett}, {Cushing}, {Marley},
  {Saumon}, {Geballe}, {Golimowski}, {Fan}, \& {Noll}}]{stephens2009}
{Stephens}, D.~C., {Leggett}, S.~K., {Cushing}, M.~C., {et~al.} 2009,
  \href{http://dx.doi.org/10.1088/0004-637X/702/1/154}{\color{magenta}\apj},
  \href{https://ui.adsabs.harvard.edu/abs/2009ApJ...702..154S}{\color{blue}702},
  \href{https://ui.adsabs.harvard.edu/abs/2009ApJ...702..154S}{\color{blue}154}

\bibitem[{{Szul{\'a}gyi} {et~al.}(2019){Szul{\'a}gyi}, {Dullemond}, {Pohl}, \&
  {Quanz}}]{Szulagyi2019}
{Szul{\'a}gyi}, J., {Dullemond}, C.~P., {Pohl}, A., \& {Quanz}, S.~P. 2019,
  \href{http://dx.doi.org/10.1093/mnras/stz1326}{\color{magenta}\mnras},
  \href{https://ui.adsabs.harvard.edu/abs/2019MNRAS.487.1248S}{\color{blue}487},
  \href{https://ui.adsabs.harvard.edu/abs/2019MNRAS.487.1248S}{\color{blue}1248}

\bibitem[{{Tanigawa} \& {Tanaka}(2016)}]{TanigawaTanaka16}
{Tanigawa}, T., \& {Tanaka}, H. 2016,
  \href{http://dx.doi.org/10.3847/0004-637X/823/1/48}{\color{magenta}\apj},
  \href{https://ui.adsabs.harvard.edu/abs/2016ApJ...823...48T}{\color{blue}823},
  \href{https://ui.adsabs.harvard.edu/abs/2016ApJ...823...48T}{\color{blue}48}

\bibitem[{{Thalmann} {et~al.}(2015){Thalmann}, {Mulders}, {Janson}, {Olofsson},
  {Benisty}, {Avenhaus}, {Quanz}, {Schmid}, {Henning}, {Buenzli}, {M{\'e}nard},
  {Carson}, {Garufi}, {Messina}, {Dominik}, {Leisenring}, {Chauvin}, \&
  {Meyer}}]{Thalmann2015}
{Thalmann}, C., {Mulders}, G.~D., {Janson}, M., {et~al.} 2015,
  \href{http://dx.doi.org/10.1088/2041-8205/808/2/L41}{\color{magenta}\apjl},
  \href{https://ui.adsabs.harvard.edu/abs/2015ApJ...808L..41T}{\color{blue}808},
  \href{https://ui.adsabs.harvard.edu/abs/2015ApJ...808L..41T}{\color{blue}L41}

\bibitem[{{Thanathibodee} {et~al.}(2019){Thanathibodee}, {Calvet}, {Bae},
  {Muzerolle}, \& {Hern{\'a}ndez}}]{Thanathibodee2019}
{Thanathibodee}, T., {Calvet}, N., {Bae}, J., {et~al.} 2019,
  \href{http://dx.doi.org/10.3847/1538-4357/ab44c1}{\color{magenta}\apj},
  \href{https://ui.adsabs.harvard.edu/abs/2019ApJ...885...94T}{\color{blue}885},
  \href{https://ui.adsabs.harvard.edu/abs/2019ApJ...885...94T}{\color{blue}94}

\bibitem[{{Vargas Catal{\'a}n} {et~al.}(2016){Vargas Catal{\'a}n}, {Huby},
  {Forsberg}, {Jolivet}, {Baudoz}, {Carlomagno}, {Delacroix}, {Habraken},
  {Mawet}, {Surdej}, {Absil}, \& {Karlsson}}]{Vargas2016}
{Vargas Catal{\'a}n}, E., {Huby}, E., {Forsberg}, P., {et~al.} 2016,
  \href{http://dx.doi.org/10.1051/0004-6361/201628739}{\color{magenta}\aap},
  \href{https://ui.adsabs.harvard.edu/abs/2016A&A...595A.127V}{\color{blue}595},
  \href{https://ui.adsabs.harvard.edu/abs/2016A&A...595A.127V}{\color{blue}A127}

\bibitem[{{Vousden} {et~al.}(2016){Vousden}, {Farr}, \& {Mandel}}]{Vousden2016}
{Vousden}, W.~D., {Farr}, W.~M., \& {Mandel}, I. 2016,
  \href{http://dx.doi.org/10.1093/mnras/stv2422}{\color{magenta}\mnras},
  \href{https://ui.adsabs.harvard.edu/abs/2016MNRAS.455.1919V}{\color{blue}455},
  \href{https://ui.adsabs.harvard.edu/abs/2016MNRAS.455.1919V}{\color{blue}1919}

\bibitem[{{Wagner} {et~al.}(2018){Wagner}, {Follete}, {Close}, {Apai}, {Gibbs},
  {Keppler}, {M{\"u}ller}, {Henning}, {Kasper}, {Wu}, {Long}, {Males},
  {Morzinski}, \& {McClure}}]{Wagner2018}
{Wagner}, K., {Follete}, K.~B., {Close}, L.~M., {et~al.} 2018,
  \href{http://dx.doi.org/10.3847/2041-8213/aad695}{\color{magenta}\apjl},
  \href{https://ui.adsabs.harvard.edu/abs/2018ApJ...863L...8W}{\color{blue}863},
  \href{https://ui.adsabs.harvard.edu/abs/2018ApJ...863L...8W}{\color{blue}L8}

\bibitem[{{Wang} {et~al.}(2015){Wang}, {Ruffio}, {De Rosa}, {Aguilar}, {Wolff},
  \& {Pueyo}}]{Wang2015}
{Wang}, J.~J., {Ruffio}, J.-B., {De Rosa}, R.~J., {et~al.} 2015, {pyKLIP: PSF
  Subtraction for Exoplanets and Disks},
  \href{http://ascl.net/1506.001}{\color{magenta}ASCL},
  \href{https://ui.adsabs.harvard.edu/abs/2015ascl.soft06001W}{\color{blue}1506.001}

\bibitem[{{Wang} {et~al.}(2018){Wang}, {Graham}, {Dawson}, {Fabrycky}, {De
  Rosa}, {Pueyo}, {Konopacky}, {Macintosh}, {Marois}, {Chiang}, {Ammons},
  {Arriaga}, {Bailey}, {Barman}, {Bulger}, {Chilcote}, {Cotten}, {Doyon},
  {Duch{\^e}ne}, {Esposito}, {Fitzgerald}, {Follette}, {Gerard}, {Goodsell},
  {Greenbaum}, {Hibon}, {Hung}, {Ingraham}, {Kalas}, {Larkin}, {Maire},
  {Marchis}, {Marley}, {Metchev}, {Millar-Blanchaer}, {Nielsen}, {Oppenheimer},
  {Palmer}, {Patience}, {Perrin}, {Poyneer}, {Rajan}, {Rameau},
  {Rantakyr{\"o}}, {Ruffio}, {Savransky}, {Schneider}, {Sivaramakrishnan},
  {Song}, {Soummer}, {Thomas}, {Wallace}, {Ward-Duong}, {Wiktorowicz}, \&
  {Wolff}}]{Wang2018}
{Wang}, J.~J., {Graham}, J.~R., {Dawson}, R., {et~al.} 2018,
  \href{http://dx.doi.org/10.3847/1538-3881/aae150}{\color{magenta}\aj},
  \href{https://ui.adsabs.harvard.edu/abs/2018AJ....156..192W}{\color{blue}156},
  \href{https://ui.adsabs.harvard.edu/abs/2018AJ....156..192W}{\color{blue}192}

\bibitem[{{Woitke} \& {Helling}(2003)}]{Woitke2003}
{Woitke}, P., \& {Helling}, C. 2003,
  \href{http://dx.doi.org/10.1051/0004-6361:20021734}{\color{magenta}\aap},
  \href{https://ui.adsabs.harvard.edu/abs/2003A&A...399..297W}{\color{blue}399},
  \href{https://ui.adsabs.harvard.edu/abs/2003A&A...399..297W}{\color{blue}297}

\bibitem[{{Woitke} \& {Helling}(2004)}]{Woitke2004}
---. 2004,
  \href{http://dx.doi.org/10.1051/0004-6361:20031605}{\color{magenta}\aap},
  \href{https://ui.adsabs.harvard.edu/abs/2004A&A...414..335W}{\color{blue}414},
  \href{https://ui.adsabs.harvard.edu/abs/2004A&A...414..335W}{\color{blue}335}

\bibitem[{{Xuan} {et~al.}(2018){Xuan}, {Mawet}, {Ngo}, {Ruane}, {Bailey},
  {Choquet}, {Absil}, {Alvarez}, {Bryan}, {Cook}, {Femen{\'\i}a Castell{\'a}},
  {Gomez Gonzalez}, {Huby}, {Knutson}, {Matthews}, {Ragland}, {Serabyn}, \&
  {Zawol}}]{Xuan2018}
{Xuan}, W.~J., {Mawet}, D., {Ngo}, H., {et~al.} 2018,
  \href{http://dx.doi.org/10.3847/1538-3881/aadae6}{\color{magenta}\aj},
  \href{https://ui.adsabs.harvard.edu/abs/2018AJ....156..156X}{\color{blue}156},
  \href{https://ui.adsabs.harvard.edu/abs/2018AJ....156..156X}{\color{blue}156}

\bibitem[{{Zhu}(2015)}]{Zhu2015}
{Zhu}, Z. 2015,
  \href{http://dx.doi.org/10.1088/0004-637X/799/1/16}{\color{magenta}\apj},
  \href{https://ui.adsabs.harvard.edu/abs/2015ApJ...799...16Z}{\color{blue}799},
  \href{https://ui.adsabs.harvard.edu/abs/2015ApJ...799...16Z}{\color{blue}16}

\end{thebibliography}

\end{CJK*}
\end{document}